%% file: main.tex
\documentclass[fleqn,vecphys,envcountchap]{svmult}

\usepackage{makeidx}          
\usepackage{graphicx}         
\usepackage{cite}             
\usepackage{multicol}         
\usepackage[bottom]{footmisc} %
\usepackage{amsmath}          %
\usepackage{amssymb}          %
\usepackage{amsfonts}         %
\usepackage{url}              %
\usepackage{bm}               %
\usepackage{mathrsfs}         %
\usepackage{feynmp}           %
\usepackage[nottoc,notbib]{tocbibind}

\makeindex                    

\calctocindent

\begin{document}

\newcommand{\beq}{\begin{equation}}
\newcommand{\eeq}{\end{equation}}
\newcommand{\bqa}{\begin{eqnarray}}
\newcommand{\eqa}{\end{eqnarray}}
\newcommand{\da}{\ensuremath{d^\dagger}}
\newcommand{\ha}{\ensuremath{h^\dagger}}
\newcommand{\adag}{\ensuremath{a^\dagger}}
\newcommand{\no}{\nonumber}
\newcommand{\ep}{\ensuremath{\epsilon}}
\newcommand{\ca}{\ensuremath{c^\dagger}}
\newcommand{\ga}{\ensuremath{\gamma^\dagger}}
\newcommand{\gm}{\ensuremath{\gamma}}
\newcommand{\up}{\ensuremath{\uparrow}}
\newcommand{\dn}{\ensuremath{\downarrow}}
\newcommand{\ms}{\medskip}
\newcommand{\bs}{\bigskip}
\newcommand{\kk}{\ensuremath{{\bf k}}}
\newcommand{\kp}{\ensuremath{{\bf k'}}}
\newcommand{\kpp}{\ensuremath{{\bf k''}}}
\newcommand{\qq}{\ensuremath{{\bf q}}}
\newcommand{\rr}{\ensuremath{{\bf r}}}
\newcommand{\rp}{\ensuremath{{\bf r'}}}
\newcommand{\nbr}{\ensuremath{\langle ij \rangle}}
\newcommand{\ncap}{\ensuremath{\hat{n}}}
\newcommand{\sigbar}{\ensuremath{\overline{\sigma}}}
\newcommand{\pr}{\ensuremath{{\cal P}}}
\newcommand{\can}{\ensuremath{{\cal S}}}

\newcommand\aSe\alpha
\newcommand\bSe\beta
\newcommand\tSe\theta
\newcommand\LSe\langle
\newcommand\rSe\right
\newcommand\lSe\left
\newcommand{\ReSe}{\,{\rm Re}\,}
\newcommand{\ImSe}{\,{\rm Im}\,}

\newcommand\hc{{\mathrm{H.c.}}}
\newcommand\Gcv{{\Gc\kern-0.76em\Gc}}
\newcommand\Gvb{{\skew4\bar\Gv}}
\newcommand\Quad{\qquad\qquad}
\newcommand\g\gamma
\newcommand\la\lambda
\newcommand\La\Lambda
\newcommand\p\varphi
\newcommand\eps\varepsilon
\newcommand\om\omega
\newcommand\Om\Omega
\newcommand\s\sigma
\newcommand{\BEQ}{\begin{equation}}
\newcommand{\EEQ}{\end{equation}}
\newcommand{\ALIGN}[1]{\begin{align}#1\end{align}}
\newcommand{\SPLIT}[1]{\begin{equation}\begin{split}#1\end{split}\end{equation}}
\newcommand{\ALIGNED}[1]{\begin{aligned}#1\end{aligned}}
\newcommand{\BEA}{\begin{eqnarray}}
\newcommand{\EEA}{\end{eqnarray}}
\newcommand\NN{\nonumber}
\newcommand\R\rangle
\newcommand\DISP\displaystyle
\newcommand\matrixp[1]{\begin{pmatrix}#1\end{pmatrix}}
\newcommand\matrixv[1]{\begin{vmatrix}#1\end{vmatrix}}
\newcommand{\BZ}{\mathrm{BZ}}
\newcommand{\G}{\Gamma}
\newcommand{\pvec}{\wedge}
\newcommand{\dg}{\dagger}
\newcommand{\mtext}[1]{\quad\text{#1}\quad}
\newcommand\tr{\,\mathrm{tr}\,}
\newcommand\Tr{\,\mathrm{Tr}\,}
\newcommand\cpp{C\raise 1.5pt\hbox{$\scriptstyle\kern-2pt++$}}
\newcommand{\Hc}{\mathscr{H}}
\newcommand{\Dc}{\mathscr{D}}
\newcommand{\Gc}{\mathscr{G}}
\newcommand{\Oc}{\mathcal{O}}
\newcommand{\Nc}{\mathscr{N}}
\newcommand{\Gg}{\mathfrak{G}}
\newcommand\ffrac[2]{{\textstyle{#1\over #2}}}
\newcommand\hf{\frac12}
\newcommand\hff{\ffrac12}
\newcommand\Frac[2]{\frac{\displaystyle #1}{\displaystyle #2}}
\newcommand\pfrac[2]{\l(\frac{#1}{#2}\r)}
\newcommand\vfrac[2]{\l|\frac{#1}{#2}\r|}
\newcommand\del[2]{\frac{\mathrm{d} #1}{\mathrm{d} #2}}
\newcommand\pdel[2]{\Frac{\partial #1}{\partial #2}}
\newcommand\ppdel[2]{\l(\pdel{#1}{#2}\r)}
\newcommand\ddel[2]{\Frac{\delta #1}{\delta #2}}
\newcommand{\varia}[2]{\frac{\delta #1}{\delta #2}}
\newcommand\er{\mathrm{e}}
\newcommand\dr{\mathrm{d}}
\newcommand\av{{\vec a}}
\newcommand\Av{{\vec A}}
\newcommand\bv{{\vec b}}
\newcommand\Bv{{\vec B}}
\newcommand\cv{{\vec c}}
\newcommand\Cv{{\vec C}}
\newcommand\dv{{\vec d}}
\newcommand\Dv{{\vec D}}
\newcommand\ev{{\vec e}}
\newcommand\Ev{{\vec E}}
\newcommand\epsv{{\vec\varepsilon}}
\newcommand\fv{{\vec f}}
\newcommand\Fv{{\vec F}}
\newcommand\Gammav{{\vec\Gamma}}
\newcommand\gv{{\vec g}}
\newcommand\Gv{{\vec G}}
\newcommand\Hv{{\vec H}}
\newcommand\id{{\vec 1}}
\newcommand\Jv{{\vec J}}
\newcommand\kv{{\vec k}}
\newcommand\Kv{{\vec K}}
\newcommand\Lambdav{{\vec\Lambda}}
\newcommand\lv{{\vec l}}
\newcommand\Lv{{\vec L}}
\newcommand\mv{{\vec m}}
\newcommand\Mv{{\vec M}}
\newcommand\nv{{\vec n}}
\newcommand\Nv{{\vec N}}
\newcommand\Ov{{\vec O}}
\newcommand\pv{{\vec p}}
\newcommand\Pv{{\vec P}}
\newcommand\qv{{\vec q}}
\newcommand\Qv{{\vec Q}}
\newcommand\rv{{\vec r}}
\newcommand\Rv{{\vec R}}
\newcommand\Sigmav{{\vec\Sigma}}
\newcommand\sv{{\vec s}}
\newcommand\Sv{{\vec S}}
\newcommand\thetav{{\vec\theta}}
\newcommand\tv{{\vec t}}
\newcommand\Tv{{\vec T}}
\newcommand\uv{{\vec u}}
\newcommand\Uv{{\vec U}}
\newcommand\vv{{\vec v}}
\newcommand\Vv{{\vec V}}
\newcommand\wv{{\vec w}}
\newcommand\Wv{{\vec W}}
\newcommand\xv{{\vec x}}
\newcommand\Xv{{\vec X}}
\newcommand\yv{{\vec y}}
\newcommand\rvt{{\tilde\rv}}
\newcommand\kvt{{\tilde\kv}}
\newcommand\qvt{{\tilde\qv}}
\newcommand\Lambdat{{\skew5\tilde\Lambdav}}
\newcommand\Qvt{{\skew3\tilde\Qv}}
\newcommand\BA{\begin{array}}
\newcommand\EA{\end{array}}
\newcommand{\MATRIX}[1]{\begin{pmatrix}#1\end{pmatrix}}

\newcommand{\be}{\begin{equation}}
\newcommand{\ee}{\end{equation}}
\newcommand{\ba}{\begin{eqnarray}}
\newcommand{\ea}{\end{eqnarray}}
\newcommand{\ff}[1]{{\bm #1}}
\newcommand{\trPo}{\mbox{tr}}
\newcommand{\TrPo}{\mbox{Tr}}
\newcommand{\refeq}[1]{Eq.\ (\ref{Potthoff:eq:#1})}
\newcommand{\labeq}[1]{\label{Potthoff:eq:#1}}
\newcommand{\reffig}[1]{Fig.\ \ref{Potthoff:fig:#1}}
\newcommand{\labfig}[1]{\label{Potthoff:fig:#1}}
\newcommand{\rem}[1]{{\color{red}#1}}
\newcommand{\e}[1]{{\color{red} #1}}
\newcommand{\bi}{\begin{itemize}}
\newcommand{\ei}{\end{itemize}}

\newcommand{\mtin}[1]{\mbox{\tiny {#1}}}

\newcommand\rJa{{\bf{r}}}
\newcommand\kJa{{\bf{k}}}
\newcommand\epJa{\epsilon}
\newcommand\GcJa{{G_c}}

\newcommand\K{{\bf{K}}}
\newcommand\tk{\tilde{\bf{k}}}
\newcommand\kt{{\tilde{\kJa}}}
\newcommand\M{{\bf M}}
\newcommand\Dk{\Delta k}
\newcommand\Sigmac{{\Sigma_c}}
\newcommand\Gammac{{\Gamma_c}}
\newcommand\Gscript{{\cal{G}}}
\newcommand\Gbar{\bar{G}}
\newcommand\si{\sigma}
\newcommand\q{{\bf{q}}}
\newcommand\chibar{\bar{\chi}}
\newcommand\chic{{\chi_c}}
\newcommand\Q{{\bf{Q}}}

\frontmatter


\mainmatter






\include{Authors/Tremblay/Tremblay}

\backmatter
\printindex
\listoffigures
\listoftables


\end{document}

%% file: Authors/Tremblay/Tremblay.tex
\title{Two-Particle-Self-Consistent Approach for the Hubbard Model}\label{Avella:Tremblay}
\author{A.-M.S. Tremblay}
\institute{D\'epartement de physique, and RQMP, Universit\'e de Sherbrooke, Sherbrooke, QC J1K 2R1, Canada
\texttt{tremblay@physique.usherbrooke.ca}
\and Canadian Institute for Advanced Research, Toronto, Ontario, Canada}
\titlerunning{Two-Particle-Self-Consistent Theory}
\toctitle{Two-Particle-Self-Consistent Theory}
\maketitle

\begin{abstract}
Even at weak to intermediate coupling, the Hubbard model poses a formidable
challenge. In two dimensions in particular, standard methods such as the
Random Phase Approximation are no longer valid since they predict a finite
temperature antiferromagnetic phase transition prohibited by the
Mermin-Wagner theorem. The Two-Particle-Self-Consistent (TPSC) approach
satisfies that theorem as well as particle conservation, the Pauli principle,
the local moment and local charge sum rules. The self-energy formula does
not assume a Migdal theorem. There is consistency between one- and
two-particle quantities. Internal accuracy checks allow one to test the
limits of validity of TPSC. Here I present a pedagogical review of TPSC
along with a short summary of existing results and two case
studies: a) the opening of a pseudogap in two dimensions when the correlation
length is larger than the thermal de Broglie wavelength, and b) the
conditions for the appearance of d-wave superconductivity in the
two-dimensional Hubbard model.
\end{abstract}

\setcounter{minitocdepth}{2}
\dominitoc

\section{Introduction}

\index{TPSC|see{Two-Particle-Self-Consistent Approach}}
\index{Two-Particle-Self-Consistent Approach}
\index{Hubbard model}
\index{self-consistency!two-particle}
Very few models can describe complex behavior observed in nature with an
economy of parameters. The Hubbard model is in this category. It has become
the cornerstone of correlated electron physics. On the down side, it is
extremely difficult to solve. While it was proposed in 1963
\cite{Tremblay:Hubbard:1963,Tremblay:Kanamori:1963,Tremblay:Gutzwiller:1963}, the only exact
results that we know are in one dimension \cite{Tremblay:Lieb:1968} and in infinite dimension \cite{Tremblay:Metzner:1989}. A
variety of approximate approaches to solve this model exist, as can be
checked from the table of contents of this volume. The
Two-Particle-Self-Consistent (TPSC) \cite{Tremblay:Vilk:1994,Tremblay:Vilk:1997,Tremblay:Mahan} approach that is described in the
present Chapter is in the category of non-perturbative semi-analytical
approaches. By semi-analytical, I mean that while it is possible to find
many analytical results, numerical integrations are necessary in the end to
obtain quantitative results.

\index{pseudogap}
\index{superconductivity}
Why should you bother to learn yet another approach? Because in its known
regime of applicability it is extremely reliable, as can be judged by
benchmark Quantum Monte Carlo (QMC) calculations. Because it satisfies a
number of exact results that control the quality of the approximation and make it physically appealing.
And because it gives physical insight into many questions related to the two-dimensional
Hubbard model relevant for the high-temperature superconductors and many
other materials. As a case study, I discuss in this Chapter the physics of
pseudogap induced by precursors to long-range order. We will see that this
describes the physics of the pseudogap in electron-doped high-temperature
superconductors where predictions of TPSC have been verified experimentally.
Pseudogap phenomena include the appearance of a minimum in the single
particle spectral weight and density of state at the Fermi level. I will be more precise later.

This Chapter offers to the reader a simple pedagogical introduction to this
approach along with the case studies mentioned above and a guide to various
other problems that have been, or have not yet been, solved with TPSC.

I assume familiarity with the basics of many-body theory, i.e. the
canonical formalism, second quantization, many-body Green's functions,
response functions and with the Matsubara formalism for finite temperature
calculations. Knowledge of functional derivative approaches would be useful
for some of the more advanced topics, but it is not essential to learn the
important results.

Before you read on, you might be interested to know a little more about the
method to decide whether it is worth the effort. TPSC is designed to study
the one-band Hubbard model
\begin{equation}
H=-\sum_{ij\sigma }t_{i,j} c_{i\sigma }^{\dagger }c_{j\sigma
} +U\sum_{i}n_{i\uparrow
}n_{i\downarrow \,\,\,\,}  \label{Tremblay:Hubbard}
\end{equation}%
where the operator $c_{i\sigma }$ destroys an electron of spin $\sigma $ at
site $i$. Its adjoint $c_{i\sigma }^{\dagger }$ creates an electron and the
number operator is defined by $n_{i\sigma }=$ $c_{i\sigma }^{\dagger
}c_{i\sigma }$. The symmetric hopping matrix $t_{i,j}$ determines the band
structure, which here can be arbitrary. The screened Coulomb interaction is
represented by the energy cost $U$ of double occupation. It is also possible
to generalize to cases where near-neighbor interactions are included. We
work in units where $k_{B}=1$, $\hbar =1$ and the lattice spacing is also
unity, $a=1$. In all numerical calculations, we take as unit of energy the
nearest-neighbor hopping $t=1.$

One of the first concepts that is discussed with the Hubbard model is that
of the Mott transition \cite{Tremblay:Mott:1949}. When dimension is larger than unity, at \textquotedblleft strong coupling\textquotedblright, large $U/t,$ the states are localized, but at \textquotedblleft weak coupling\textquotedblright, small $U/t$, the states are delocalized. The Mott transition is quite subtle and has been the subject of many papers. It is discussed in the Chapters of M. Jarrell, M. Potthoff, D. S\'en\'echal  and others in this volume.

\index{Pauli principle}
TPSC is valid from weak to intermediate coupling. Hence, on the negative side, it
does not describe the Mott transition.
Nevertheless, there is a large number
of physical phenomena that it allows to study. An important one is
antiferromagnetic fluctuations in two- or higher-dimensional lattices. A standard Random Phase Approximation (RPA) calculation of the spin
susceptibility signals a finite temperature phase transition to
antiferromagnetic long-range order. This is prohibited by the Mermin-Wagner
theorem \cite{Tremblay:Mermin:1966,Tremblay:Hohenberg:1966} that states that in two dimensions you cannot break a continuous
symmetry at finite temperature. It is extremely important physically that in two dimensions there is
a wide range of temperatures where there are huge antiferromagnetic
fluctuations in the paramagnetic state. The standard way to treat fluctuations
in many-body theory, RPA misses this.
As we will see, the RPA also violates the Pauli principle in an
important way. The composite operator method (COM), described in this volume by Avella and Mancini (see Chap.~\ref{Avella:Mancini}), is another approach that satisfies the Mermin-Wagner theorem and the Pauli principle \cite{Tremblay:Mancini:2004,Tremblay:Mancini:2006,Tremblay:Mancini:2007}. What other  approaches satisfy the Mermin-Wagner theorem at weak coupling? The Fluctuation Exchange Approximation (FLEX) \index{fluctuation exchange approximation} \cite{Tremblay:Bickers:1989,Tremblay:Bickers_dwave:1989}, and the
self-consistent renormalized theory of Moriya-Lonzarich \cite{Tremblay:Moriya:2003,Tremblay:Lonzarich:1985,Tremblay:Moriya:1985}.
Each has its strengths and
weaknesses, as discussed in Refs. \cite{Tremblay:Vilk:1997,Tremblay:Allen:2003}. Weak coupling renormalization group approaches \footnote{See the contribution of C. Honerkamp in this volume.} become uncontrolled
when the antiferromagnetic fluctuations begin to diverge \cite{Tremblay:Dzyaloshinskii:1987,Tremblay:Schulz:1987,Tremblay:Lederer:1987,Tremblay:Honerkamp:2001}. Other approaches include
the effective spin-Hamiltonian approach \cite{Tremblay:Logan:1996}.

In summary, the advantages and disadvantages of TPSC are as follows.
Advantages:

\index{Mermin-Wagner-Hohenberg theorem}
\begin{itemize}
\item There are no adjustable parameters.

\item Several exact results are satisfied: Conservation laws for spin and
charge, the Mermin-Wagner theorem, the Pauli principle in the form $%
\left\langle n_{\uparrow }^{2}\right\rangle =\left\langle n_{\uparrow
}\right\rangle,$ the local moment and local-charge sum rules and the f
sum-rule.

\item Consistency between one and two-particle properties serves as a
guide to the domain of validity of the approach. (Double occupancy
obtained from sum rules on spin and charge equals that obtained from the self-energy and the Green
function).

\item Up to intermediate coupling, TPSC agrees within a few percent with
Quantum Monte Carlo (QMC) calculations. Note that QMC calculations can serve
as benchmarks since they are exact within statistical accuracy, but they are
limited in the range of physical parameter accessible because of the sign problem.

\item We do not need to assume that Migdal's theorem applies to be able to
obtain the self-energy.
\end{itemize}

The main successes of TPSC that I will discuss include

\begin{itemize}
\item Understanding the physics of the pseudogap induced by precursors of a
long-range ordered phase in two dimensions. For this understanding, one
needs a method that satisfies the Mermin-Wagner theorem to create a broad
temperature range where the antiferromagnetic correlation length is larger
than the thermal de Broglie wavelength. That method must also allow one to
compute the self-energy reliably. Only TPSC does both.

\item Explaining the pseudogap in electron-doped cuprate superconductors
over a wide range of dopings.

\item Finding estimates of the transition temperature for d-wave
superconductivity that were found later in agreement with quantum cluster
approaches such as the Dynamical Cluster Approximation.

\item Giving quantitative estimates of the range of temperature where
quantum critical behavior can affect the physics.
\end{itemize}

The drawbacks of this approach, that I explain as we go along, are that

\begin{itemize}
\item It works well in two or more dimensions, not in one dimension \footnote{Modifications have been proposed in zero dimension to use as impurity solver for DMFT \cite{Tremblay:Georges:1996}} \cite{Tremblay:Nelisse:1999}.

\item It is not valid at strong coupling, except at very high temperature
where it recovers the atomic limit \cite{Tremblay:Dare:2007}.

\item It is not valid deep in the renormalized classical regime \cite{Tremblay:Vilk:1994}.

\item For models other than the one-band Hubbard model, one usually runs out
of sum rules and it is in general not possible to find all parameters
self-consistently. With nearest-neighbor repulsion, it has been possible to
find a way out as I will discuss below.
\end{itemize}

\index{quantum Monte Carlo method!and TPSC}
\index{superconductivity}
For detailed comparisons with QMC calculations, discussions of the physics
and detailed comparisons with other approaches, you can refer to Ref.\cite{Tremblay:Vilk:1997,Tremblay:Allen:2003}. You
can read Ref.\cite{Tremblay:LTP:2006} for a review of the work related to the pseudogap and
superconductivity up to 2005 including detailed comparisons with Quantum
Cluster approaches in the regime of validity that overlaps with TPSC
(intermediate coupling).

Section \ref{Tremblay:Method} introduces TPSC in the simplest physically motivated
way and demonstrates the various results that are exactly satisfied. The
following section \ref{Tremblay:Case} presents two case studies: the pseudogap and
d-wave superconductivity. Many more known results and extensions are
summarized in section \ref{Tremblay:Repulsive}. The attractive Hubbard model is in
the next to last section \ref{Tremblay:Attractive Hubbard}. I conclude with some
open problems in section \ref{Tremblay:Open}.

\section{The method\label{Tremblay:Method}}

In the first part of this section, I present TPSC as if we were discussing
in front of a chalkboard. More formal ways of presenting the results come
later.

\subsection{Physically motivated approach, spin and charge fluctuations\label{Tremblay:Physically-motivated}}

\index{Two-Particle-Self-Consistent Approach!physical motivation}
As basic physical requirements, we would like our approach to satisfy a)
conservation laws, b) the Pauli principle and c) the Mermin-Wagner-Hohenberg-Coleman theorem.
The standard RPA approach satisfies the first requirement but not the other
two. Let us see this. With the charge and spin given by%
\begin{equation}
n_{i}\equiv n_{i\uparrow }+n_{i\downarrow }
\end{equation}%
\begin{equation}
S_{i}^{z}\equiv n_{i\uparrow }\left( \tau \right) -n_{i\downarrow }\left(
\tau \right) .
\end{equation}%
the RPA spin and charge susceptibilities in the one-band Hubbard model are given
respectively by%
\begin{equation}
\chi _{sp}(q)=\frac{\chi _{0}(q)}{1-\frac{1}{2}U\chi _{0}(q)}\;;\;\chi
_{ch}(q)=\frac{\chi _{0}(q)}{1+\frac{1}{2}U\chi _{0}(q)}
\end{equation}%
with $q$ a short-hand for both wave vector $\mathbf{q}$ and
Matsubara frequency and where $\chi _{0}(q)$ is the Lindhard function that in analytically continued
retarded form is, for a discrete lattice of $N$ sites,
\begin{equation}
\chi ^{0R}(\mathbf{q},\omega )=-\frac{2}{N}\sum_{\mathbf{k}}\frac{%
f\left( \varepsilon _{\mathbf{k}}\right) -f\left( \varepsilon _{\mathbf{k+q}%
}\right) }{\omega +i\eta +\varepsilon _{\mathbf{k}}-\varepsilon _{\mathbf{k+q%
}}}.  \label{Tremblay:Lindhard}
\end{equation}%
In this expression, assuming periodic boundary conditions,%
\begin{equation}
\varepsilon _{\mathbf{k}}=\left( -\sum_{j}e^{i\mathbf{k\cdot }\left( \mathbf{%
r}_{i}-\mathbf{r}_{j}\right) }t_{i,j}\right)-\mu
\end{equation}%
with the sum over $j$ is running over all neighbors of any of the sites $i.$
The chemical potential $\mu $ is chosen so that we have the required density.

It is known on general grounds \cite{Tremblay:Baym:1962} that RPA satisfies conservation
laws, but it is easy to check that for a special case. Since spin and charge
are conserved, then the equalities $\chi _{sp}^{R}(\mathbf{q=0,}\omega )=0$ and $\chi
_{ch}^{R}(\mathbf{q=0,}\omega )=0$ for $\omega \neq 0$ follow from the
corresponding equality for the non-interacting Lindhard function $\chi ^{0R}(%
\mathbf{q=0,}\omega )=0.$

\index{Mermin-Wagner-Hohenberg theorem!violation by RPA}
To check that RPA violates the Mermin Wagner theorem, it suffices to
note that if $U$ is larger than $U_{c}=2/\chi _{0}^{R}(\mathbf{q}_{\max
},\omega =0)$, then the denominator $1-\frac{1}{2}U\chi _{0}(q)$ of $\chi
_{sp}(q)$ can diverge at some wave vector $\mathbf{q}_{\max}$ and temperature.

\index{Pauli principle!violation by RPA}
The violation of
the Pauli principle requires a bit more thinking. We derive a sum rule that
rests on the use of the Pauli principle and check that it is violated by RPA
to second order in $U$. First note that if we sum the spin and charge susceptibilities over all wave
vectors $\mathbf{q}$ and all Matsubara frequencies $i\omega _{n}$,\footnote{In other references we often use $iq_n$ instead of $i\omega_n$ to denote the Matsubara
frequency corresponding to wave vector $\mathbf{q}$.} we obtain local, equal-time correlation
functions, namely%
\begin{equation}
\frac{T}{N}\sum_{\mathbf{q}}\sum_{i\omega _{n}}\chi _{sp}(\mathbf{q,}i\omega
_{n})=\left\langle \left( n_{\uparrow }-n_{\downarrow }\right)
^{2}\right\rangle =\left\langle n_{\uparrow }\right\rangle +\left\langle
n_{\downarrow }\right\rangle -2\left\langle n_{\uparrow }n_{\downarrow
}\right\rangle  \label{Tremblay:sumSpin}
\end{equation}%
and
\begin{equation}
\frac{T}{N}\sum_{\mathbf{q}}\sum_{i\omega _{n}}\chi _{ch}(\mathbf{q,}i\omega
_{n})=\left\langle \left( n_{\uparrow }+n_{\downarrow }\right)
^{2}\right\rangle -\left\langle n_{\uparrow }+n_{\downarrow }\right\rangle
^{2}=\left\langle n_{\uparrow }\right\rangle +\left\langle n_{\downarrow
}\right\rangle +2\left\langle n_{\uparrow }n_{\downarrow }\right\rangle
-n^{2}  \label{Tremblay:sumCharge}
\end{equation}%
where on the right-hand side, we used the Pauli principle $n_{\sigma
}^{2}=\left( c_{\sigma }^{\dagger }c_{\sigma }\right) \left( c_{\sigma
}^{\dagger }c_{\sigma }\right) =c_{\sigma }^{\dagger }c_{\sigma }-c_{\sigma
}^{\dagger }c_{\sigma }^{\dagger }c_{\sigma }c_{\sigma }=c_{\sigma
}^{\dagger }c_{\sigma }=n_{\sigma }$ that follows from $c_{\sigma }^{\dagger
}c_{\sigma }^{\dagger }=c_{\sigma }c_{\sigma }=0.$ This is the simplest
version of the Pauli principle. Full antisymmetry is another matter \cite{Tremblay:Bickers:1991,Tremblay:Janis:2006}. We call
the first of the above displayed equations the local spin sum-rule and the
second one the local charge sum-rule. For RPA, adding the two sum rules
yields \index{Pauli principle!as a sum rule}
\begin{eqnarray}
\frac{T}{N}\sum_{\mathbf{q}}\sum_{i\omega _{n}}\left( \chi _{sp}(\mathbf{q,}%
i\omega _{n})+\chi _{ch}(\mathbf{q,}i\omega _{n})\right) &=& \\
\frac{T}{N}\sum_{q}\left( \frac{\chi _{0}(q)}{1-\frac{1}{2}U\chi _{0}(q)}\;+%
\frac{\chi _{0}(q)}{1+\frac{1}{2}U\chi _{0}(q)}\right) &=&2n-n^{2}.
\label{Tremblay:Pauli_sum}
\end{eqnarray}%
Since the non-interacting susceptibility $\chi _{0}(q)$ satisfies the sum
rule, we see by expanding the denominators that in the interacting case it
is violated already to second order in $U$ because $\chi _{0}(q)$ being real
and positive, (See Eq.(\ref{Tremblay:Spectral_rep_chi})), the quantity $\sum_{q}\chi _{0}(q)^{3}$ cannot vanish.

\index{Pauli principle!verified by TPSC}
How can we go about curing this violation of the Pauli principle while not
damaging the conserving aspects? The simplest way is to proceed in the
spirit of Fermi liquid theory and assume that the effective interaction
(irreducible vertex in the jargon) is renormalized. This renormalization has
to be different for spin and charge so that%
\begin{eqnarray}
\chi _{sp}(q) &=&\frac{\chi ^{\left( 1\right) }(q)}{1-\frac{1}{2}U_{sp}\chi
^{\left( 1\right) }(q)}\;\;  \label{Tremblay:chi_sp} \\
\chi _{ch}(q) &=&\frac{\chi ^{\left( 1\right) }(q)}{1+\frac{1}{2}U_{ch}\chi
^{\left( 1\right) }(q)}.  \label{Tremblay:chi_ch}
\end{eqnarray}%
In practice $\chi ^{\left( 1\right) }(q)$ is the same\footnote{%
The meaning of the superscripts differs from that in Ref. \cite{Tremblay:Vilk:1997}.
Superscripts $\left( 2\right) \left( 1\right) $ here correspond respectively
to $\left( 1\right) \left( 0\right) $ in Ref. \cite{Tremblay:Vilk:1997}} as the Lindhard
function $\chi _{0}(q)$ Eq.(\ref{Tremblay:Lindhard}) for $U=0$ but, strictly speaking, there is a
constant self-energy term that is absorbed in the definition of $\mu $
\cite{Tremblay:Allen:2003}. We are almost done with the collective modes. Substituting the
above expressions for $\chi _{sp}(q)$ and $\chi _{ch}(q)$ in the two
sum-rules, local-spin and local-charge appearing in Eqs.(\ref{Tremblay:sumSpin},\ref{Tremblay:sumCharge}), we could determine both $U_{sp}$ and $U_{ch}$ if we knew $%
\left\langle n_{\uparrow }n_{\downarrow }\right\rangle .$ The following
\textit{ansatz }%
\begin{equation}
U_{sp}\left\langle n_{\uparrow }\right\rangle \left\langle n_{\downarrow
}\right\rangle =U\left\langle n_{\uparrow }n_{\downarrow }\right\rangle
\label{Tremblay:ansatz}
\end{equation}%
gives us the missing equation. Now notice that $U_{sp},$ or equivalently $%
\left\langle n_{\uparrow }n_{\downarrow }\right\rangle $ depending on which
of these variables you want to treat as independent, is determined
self-consistently. That explains the name of the approach,
``Two-Particle-Self-Consistent''. Since the the sum-rules are satisfied
exactly, when we add them up the resulting equation, and hence the Pauli
principle, will also be satisfied exactly. In other words, in Eq.(\ref{Tremblay:Pauli_sum}) that follows from the Pauli principle, we now have $U_{sp}$ and
$U_{ch}$ on the left-hand side that arrange each other in such a way that
there is no violation of the principle. In standard many-body theory, the Pauli principle (crossing symmetry) is achieved in a much more complicated way by solving parquet equations. \cite{Tremblay:Janis:2006,Tremblay:Bickers:1991}

The ansatz Eq.(\ref{Tremblay:ansatz}) is inspired from the work of Singwi \cite{Tremblay:Singwi:1981,Tremblay:Ichimaru:1982} and
was also found independently by M. R. Hedeyati and G. Vignale \cite{Tremblay:Hedeyati:1989}.
The whole procedure is not as arbitrary as it may seem and
we justify this in more detail in section \ref{Tremblay:Formal} with the formal derivation.
For now, let us just add a few physical considerations.

\begin{remark}
Since $U_{sp}$ and $U_{ch}$ are renormalized with respect to the bare value,
one might have expected that one should use the dressed Green's
functions in the calculation of $\chi _{0}\left( q\right).$ It is
explained in appendix A of Ref.\cite{Tremblay:Vilk:1997} that this would lead to a
violation of the results $\chi _{sp}^{R}(\mathbf{q=0,}\omega )=0$ and $\chi
_{ch}^{R}(\mathbf{q=0,}\omega )=0$. In the present approach, the f-sum rule
\begin{eqnarray}
\int \frac{d\omega }{\pi }\omega \chi _{ch,sp}^{\prime \prime }\left(
\mathbf{q,}\omega \right)  &=&\lim_{\eta \rightarrow 0}T\sum_{i\omega
_{n}}\left( e^{-i\omega _{n}\eta }-e^{i\omega _{n}\eta }\right) i\omega
_{n}\chi _{ch,sp}\left( \mathbf{q},i\omega _{n}\right)  \\
&=&\frac{1}{N}\sum_{\mathbf{k}\sigma }\left( \epsilon _{\mathbf{k+q}%
}+\epsilon _{\mathbf{k-q}}-2\epsilon _{\mathbf{k}}\right) n_{\mathbf{k}%
\sigma }  \label{Tremblay:f-sum q dep}
\end{eqnarray}%
is satisfied with $n_{\mathbf{k}\sigma }=n_{\mathbf{k\sigma }}^{\left( 1\right) }$, which is the same as the Fermi
function for the non-interacting case since it is computed from $G^{1}$. \footnote{For the conductivity with vertex corrections \cite{Tremblay:Bergeron:2011}, the f-sum rule with $n_{\mathbf{k}\sigma }$ obtained from $G^{(2)}$ is satisfied.}
\end{remark}

\index{Two-Particle-Self-Consistent Approach!and Singwi}
\index{Pauli principle}
\begin{remark}
$U_{sp}\left\langle n_{\uparrow }\right\rangle \left\langle n_{\downarrow
}\right\rangle =U\left\langle n_{\uparrow }n_{\downarrow }\right\rangle $
can be understood as correcting the Hartree-Fock factorization to obtain the
correct double occupancy. Expressing the irreducible vertex in
terms of an equal-time correlation function is inspired by the approach of
Singwi \cite{Tremblay:Singwi:1981} to the electron gas. But TPSC is different since it also
enforces the Pauli principle and connects to a local correlation function,
namely $\left\langle n_{\uparrow }n_{\downarrow }\right\rangle .$
\end{remark}

\subsection{Mermin-Wagner, Kanamori-Brueckner and benchmarking spin and
charge fluctuations \label{Tremblay:Mermin Wagner}}

\index{Mermin-Wagner-Hohenberg theorem}
The results that we found for spin and charge
fluctuations have the RPA form but the renormalized interactions $U_{sp}$
and $U_{ch}$ must be computed from%
\begin{equation}
\frac{T}{N}\sum_{\mathbf{q}}\sum_{i\omega _{n}}\frac{\chi ^{\left( 1\right)
}(q)}{1-\frac{1}{2}U_{sp}\chi ^{\left( 1\right) }(q)}=n-2\left\langle
n_{\uparrow }n_{\downarrow }\right\rangle  \label{Tremblay:TPSC_spin}
\end{equation}%
and
\begin{equation}
\frac{T}{N}\sum_{\mathbf{q}}\sum_{i\omega _{n}}\frac{\chi ^{\left( 1\right)
}(q)}{1+\frac{1}{2}U_{ch}\chi ^{\left( 1\right) }(q)}=n+2\left\langle
n_{\uparrow }n_{\downarrow }\right\rangle -n^{2}.  \label{Tremblay:TPSC_charge}
\end{equation}%
With the \textit{ansatz} Eq.(\ref{Tremblay:ansatz}), the above system of equations is
closed and the Pauli principle is enforced. The first of the above equations
is solved self-consistently with the $U_{sp}$ ansatz. This gives the double
occupancy $\left\langle n_{\uparrow }n_{\downarrow }\right\rangle $ that is
then used to obtain $U_{ch}$ from the next equation. The fastest way to
numerically compute $\chi ^{\left( 1\right) }(q)$ is to use fast Fourier
transforms \cite{Tremblay:Bergeron:2011}.

These TPSC expressions for spin and charge fluctuations were obtained by
enforcing the conservations laws and the Pauli principle. In particular,
TPSC satisfies the
f-sum rule Eq.(\ref{Tremblay:f-sum q dep}).
But we obtain for free a lot more, namely Kanamori-Brueckner renormalization and the
Mermin-Wagner theorem.

\index{Kanamori renormalization of $U$}
\index{Brueckner renormalization of $U$}
Let us begin with Kanamori-Brueckner renormalization of $U$. Many years
ago, Kanamori in the context of the Hubbard model \cite{Tremblay:Kanamori:1963}, and Brueckner in
the context of nuclear physics, introduced the notion that the bare $U$
corresponds to computing the scattering of particles in the first Born
approximation. In reality, we should use the full scattering cross section
and the effective $U$ should be smaller. From Kanamori's point of view, the
two-body wave function can minimize the effect of $U$ by becoming smaller to
reduce the value of the probability that two electrons are on the same site.
The maximum energy that this can cost is the bandwidth since that is the
energy difference between a one-body wave function with no nodes and one
with the maximum allowed number. Let us see how this physics comes out of
our results. Far from phase transitions, we can expand the denominator of
the local moment sum-rule equation to obtain%
\begin{equation}
\frac{T}{N}\sum_{\mathbf{q}}\sum_{i\omega _{n}}\chi ^{\left( 1\right)
}(q)\left( 1+\frac{1}{2}U_{sp}\chi ^{\left( 1\right) }(q)\right) =n-2\frac{%
U_{sp}}{U}\left\langle n_{\uparrow }\right\rangle \left\langle n_{\downarrow
}\right\rangle .
\end{equation}%
Since
$\frac{T}{N}\sum_{\mathbf{q}}\sum_{i\omega _{n}}\chi ^{\left( 1\right)}(q)=n-2\left\langle n_{\uparrow }\right\rangle \left\langle n_{\downarrow }\right\rangle $,
we can solve for $U_{sp}$ and obtain
\footnote{There is
a misprint of a factor of 2 in Ref. \cite{Tremblay:Vilk:1997}. It is corrected in Ref.\cite{Tremblay:Dare:2007}.}
\begin{eqnarray}
U_{sp} &=&\frac{U}{1+\Lambda U} \\
\Lambda &\equiv &\frac{1}{n^{2}}\frac{T}{N}\sum_{i\omega _{n}}\sum_{\mathbf{q%
}}\left( \chi ^{\left( 1\right) }\right) ^{2}\left( \mathbf{q,}i\omega
_{n}\right).
\end{eqnarray}%
\index{particle-hole channel}
\index{particle-particle channel}
We see that at large $U,$ $U_{sp}$ saturates to $1/\Lambda $, which in
practice we find to be of the order of the bandwidth. For those that are
familiar with diagrams, note that the Kanamori-Brueckner physics amounts to
replacing each of the interactions $U$ in the ladder or bubble sum for
diagrams in the particle-hole channel by infinite ladder sums in the
particle-particle channel \cite{Tremblay:Chen:1991}. This is not quite what we obtain here
since $\left( \chi ^{\left( 1\right) }\right) ^{2}$ is in the particle-hole
channel, but in the end, numerically, the results are close and the Physics
seems to be the same. One cannot make strict comparisons between TPSC and
diagrams since TPSC is non-perturbative.

While Kanamori-Brueckner renormalization, or screening, is a quantum
effect that occurs even far from phase transitions, when we are close
we need to worry about the Mermin-Wagner theorem. To satisfy this
theorem, approximate theories must prevent $\left\langle
S_z^2\right\rangle $ from becoming infinite, which is equivalent to stoping $\left\langle
n_{\uparrow }n_{\downarrow }\right\rangle $ from taking unphysical values.
This quantity is positive and bounded by its value for $U=\infty $ and its
value for non-interacting systems, namely $0\leq \left\langle n_{\uparrow
}n_{\downarrow }\right\rangle \leq n^{2}/4$. Hence, the right-hand side of
the local-moment sum-rule Eq.(\ref{Tremblay:TPSC_spin}) is contained in the interval $%
\left[ n,n-\frac{1}{2}n^{2}\right] .$ To see how the Mermin-Wagner theorem
is satisfied, write the self-consistency condition Eq.(\ref{Tremblay:TPSC_spin}) in
the form
\begin{equation}
\frac{T}{N}\sum_{q}\frac{\chi ^{\left( 1\right) }(q)}{1-\frac{1}{2}U\frac{%
\langle n_{\uparrow }n_{\downarrow }\rangle }{\langle n_{\uparrow }\rangle
\left\langle n_{\downarrow }\right\rangle }\chi ^{\left( 1\right) }\left(
q\right) }=n-2\langle n_{\uparrow }n_{\downarrow }\rangle .  \label{Tremblay:Spin2}
\end{equation}%
Consider increasing $\langle n_{\uparrow }n_{\downarrow }\rangle $ on the
left-hand side of this equation. The denominator becomes smaller, hence the
integral larger. To become larger, $\langle n_{\uparrow }n_{\downarrow
}\rangle $ has to decrease on the right-hand side. There is thus negative
feedback in this equation that will make the self-consistent solution
finite. This, however, does not prevent the expected phase transition in three
dimensions \cite{Tremblay:Dare:1996}. To see this, we need to look in
more details at the phase space for the integral in the sum rule.

As we know from the spectral representation for $\chi ,$
\begin{equation}
\chi _{ch,sp}\left( \mathbf{q},i\omega _{n}\right) =\int \frac{d\omega
^{\prime }}{\pi }\frac{\chi _{ch,sp}^{\prime \prime }\left( \mathbf{q,}%
\omega ^{\prime }\right) }{\omega ^{\prime }-i\omega _{n}}=\int \frac{%
d\omega ^{\prime }}{\pi }\frac{\omega ^{\prime }\chi _{ch,sp}^{\prime \prime
}\left( \mathbf{q,}\omega ^{\prime }\right) }{\left( \omega ^{\prime
}\right) ^{2}+\left( \omega _{n}\right) ^{2}}.  \label{Tremblay:Spectral_rep_chi}
\end{equation}
the zero Matsubara frequency contribution is always the largest. There, we
find the so-called Ornstein-Zernicke form for the susceptibility.

\index{Ornstein-Zernicke}
\index{renormalized classical regime}
\begin{description}
\item[Ornstein-Zernicke form] Let us expand the denominator near the point where $1-\frac{1}{2}%
U_{sp}\chi ^{\left( 1\right) }(\mathbf{Q,}0)=0.$ The wave vector $\mathbf{Q}$
is that where $\chi ^{\left( 1\right) }$ is maximum. We find \cite{Tremblay:Dare:1996},
\begin{eqnarray}
\chi _{sp}\left( \mathbf{q+Q},\omega +i\eta\right) &\simeq &\frac{\chi
^{\left( 1\right) }(\mathbf{Q,}0)}{1-\frac{1}{2}U_{sp}\chi ^{\left( 1\right)
}
-\frac{1}{4}U_{sp}\frac{\partial ^{2}\chi ^{\left( 1\right) }
}{\partial \mathbf{Q}^{2}}q^{2}-\frac{1}{2}U_{sp}\frac{\partial
\chi ^{\left( 1\right) }
}{\partial  \omega }\omega}  \notag \\
&\sim &\frac{\xi ^{2}}{1+\xi ^{2}q^{2}-i\omega /\omega _{sp}},
\label{Tremblay:Chi_sp_low_frequency}
\end{eqnarray}%
where all functions and derivatives in the denominator are evaluated at
$(\mathbf{Q},0)$
and where, on dimensional grounds,
\begin{equation}
-\frac{1}{4}U_{sp}\frac{\partial ^{2}\chi
^{\left( 1\right) }(\mathbf{Q},0)}{\partial \mathbf{Q}^{2}}/\left( 1-\frac{1%
}{2}U_{sp}\chi ^{\left( 1\right) }(\mathbf{Q},0)\right)
\end{equation}
scales (noted $%
\sim $) as the square of a length, $\xi $, the correlation length. That
length is determined self-consistently. Since $\omega _{sp}\sim \xi ^{-2},$
all finite Matsubara frequency contributions are negligible if $2\pi
T/\omega _{sp}\sim 2\pi T\xi ^{2}\gg 1$. That condition in the form $\omega
_{sp}\ll T$ justifies the name of the regime we are interested in, namely
the renormalized classical regime. The classical regime of a harmonic
oscillator occurs when $\omega \ll T.$ The regime here is \textquotedblleft
renormalized\textquotedblright\ classical because at temperatures above the
degeneracy temperature, the system is a free classical gas. As temperature
decreases below the Fermi energy, it becomes quantum mechanical, then close
to the phase transition, it becomes classical again.
\end{description}

Substituting the Ornstein-Zernicke form for the zero Matsubara frequency susceptibility in the
self-consistency relation Eq.(\ref{Tremblay:TPSC_spin}), we obtain
\begin{equation}
T\int \frac{d^{d}\mathbf{q}}{\left( 2\pi \right) ^{d}}\frac{1}{q^{2}+\xi
^{-2}}=\widetilde{C}  \label{Tremblay:IntegLorentz=C}
\end{equation}%
where $\widetilde{C}$ contains all non-zero Matsubara frequency contributions as
well as $n-2\left\langle n_{\uparrow }n_{\downarrow }\right\rangle .$ Since $%
\widetilde{C}$ is finite, this means that in two dimensions $\left(
d=2\right) $, it is impossible to have $\xi ^{-2}=0$ on the left-hand side
otherwise the integral would diverge logarithmically. This is clearly a
dimension-dependent statement that proves the Mermin-Wagner theorem. In
two-dimensions, we see that the integral gives a logarithm that leads to
\begin{equation*}
\xi \sim \exp \left( C^{\prime }/T\right) .
\end{equation*}%
where in general, $C^{\prime }$ can be temperature dependent \cite{Tremblay:Dare:1996}.
When $C\,^{\prime }$ is not temperature dependent, the above result is
similar to what is found at strong coupling in the non-linear sigma model.
The above dimensional analysis is a bit expeditive. A more careful analysis
\cite{Tremblay:Moriya:1990,Tremblay:Roy:2008} yields prefactors in the temperature dependence of the
correlation length.

\begin{figure}[t]
\centering  \includegraphics*[width=.7\textwidth]{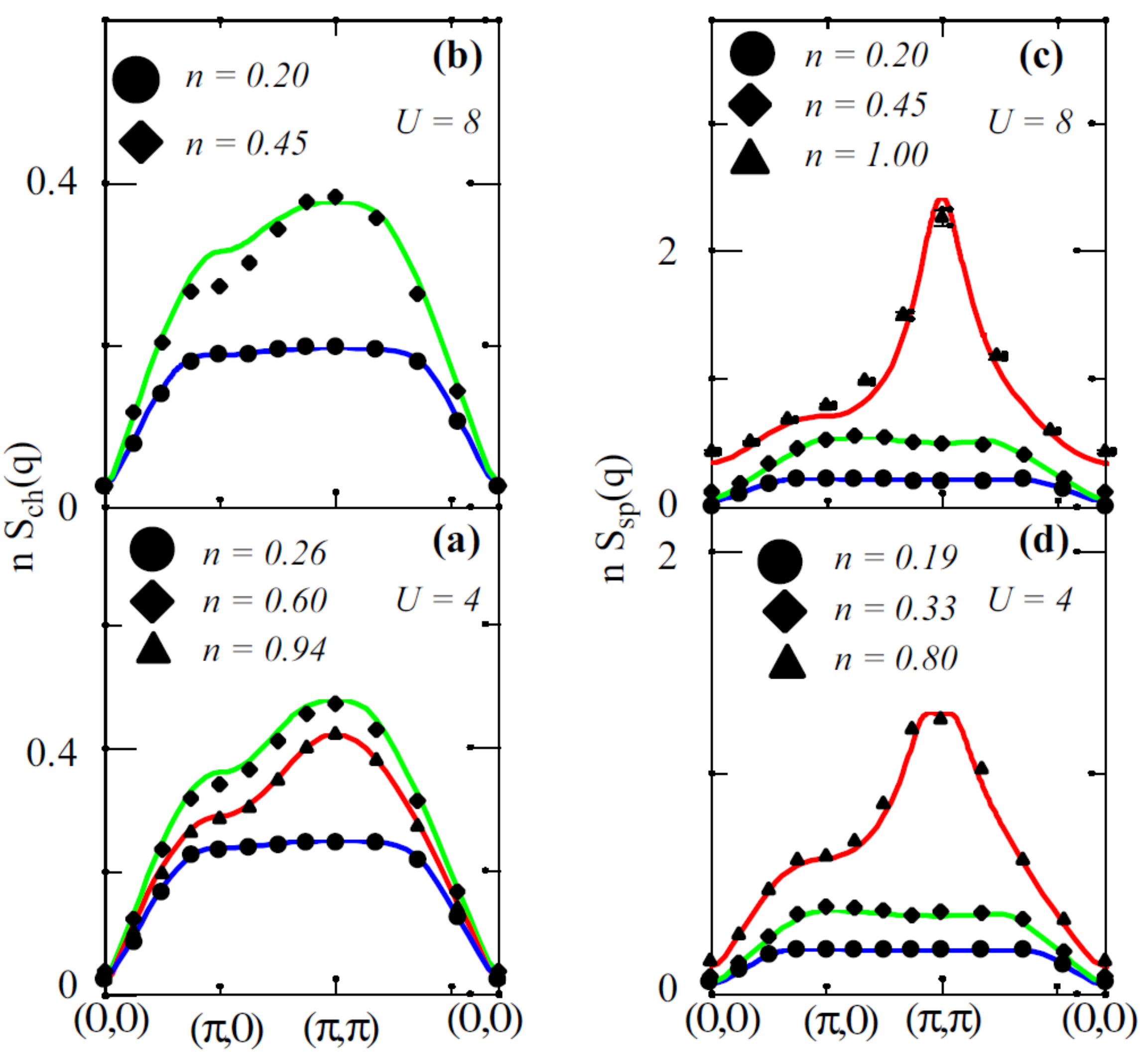}
\caption[Comparisons between TPSC and QMC for spin and charge structure factor]{Wave vector ($\mathbf{q}$) dependence of the spin and charge
structure factors for different sets of parameters. Solid lines are from
TPSC and symbols are QMC data. Monte Carlo data for $n=1$ and $U=8t$ are
for $6\times 6$ clusters and $T=0.5t$; all other data are for $8\times 8$
clusters and $T=0.2t$. Error bars are shown only when significant. From Ref.~%
\protect\cite{Tremblay:Vilk:1994}.}
\label{Tremblay:f26}
\end{figure}

\index{spin structure factor}
\index{charge structure factor}
The set of TPSC equations for spin and charge fluctuations Eqs.(\ref{Tremblay:TPSC_spin},\ref{Tremblay:TPSC_charge},\ref{Tremblay:ansatz}) is rather intuitive and simple.
The agreement of calculations with benchmark QMC calculations is rather
spectacular, as shown in Fig.(\ref{Tremblay:f26}). There, one can see the results of
QMC calculations of the structure factors, i.e. the Fourier transform of the
equal-time charge and spin correlation functions, compared with the
corresponding TPSC results.
\index{Pauli principle!influence on charge fluctuations}
This figure allows one to watch the Pauli principle in action. At $U=4t,$
Fig.(\ref{Tremblay:f26}a) shows that the charge structure factor does not have a
monotonic dependence on density. This is because, as we approach
half-filling, the spin fluctuations are becoming so large that the charge
fluctuations have to decrease so that the sum still satisfies the Pauli
principle, as expressed by Eq.(\ref{Tremblay:Pauli_sum}).

More comparisons may be found in Refs. \cite{Tremblay:LTP:2006} and \cite{Tremblay:Vilk:1994,Tremblay:Vilk:1997,Tremblay:Vilk:1996,Tremblay:Kyung:2003}.
This kind of agreement is found even at couplings of the order of the bandwidth and when second-neighbor
hopping $t^{\prime }$ is present \cite{Tremblay:Veilleux:1994,Tremblay:Veilleux:1995}.

\index{renormalized classical regime}
\begin{remark}
Even though the entry in the renormalized classical regime is well described
by TPSC \cite{Tremblay:Kyung:2003a}, equation (\ref{Tremblay:ansatz}) for $U_{sp}$ fails
deep in that regime because $\Sigma ^{\left( 1\right) }$ becomes too
different from the true self-energy. At $n=1$, $t^{\prime }=0$, deep
in the renormalized classical regime, $U_{sp}$ becomes arbitrarily small,
which is clearly unphysical. However, by assuming that $\langle n_{\uparrow
}n_{\downarrow }\rangle $ is temperature independent below $T_{X},$ a
property that can be verified from QMC calculations, one obtains a
qualitatively correct description of the renormalized-classical regime. One
can even drop the ansatz and take $\langle n_{\uparrow }n_{\downarrow
}\rangle $ from QMC on the right-hand side of the local moment sum-rule Eq.(%
\ref{Tremblay:TPSC_spin}) to obtain $U_{sp}.$
\end{remark}

\subsection{Self-energy\label{Tremblay:Sec-Self}}

\index{self-energy}
Collective charge and spin excitations can be obtained accurately from
Green's functions that contain a simple self-energy, as we have just seen.
Such modes are determined more by conservations laws than by details of the
self-energy, especially at finite temperature where the lowest fermionic
Matsubara frequency is not zero. The self-energy on the other hand is much
more sensitive to collective modes since these are important at low
frequency. The second step of TPSC is thus to find a better approximation
for the self-energy. This is similar in spirit to what is done in the
electron gas \cite{Tremblay:Mahan} where plasmons are found with non-interacting
particles and then used to compute an improved approximation for the
self-energy. This two step process is also analogous to renormalization
group calculations where renormalized interactions are evaluated to one-loop
order and quasiparticle renormalization appears only to two-loop order
\cite{Tremblay:Meyhnard:1973,Tremblay:Bourbonnais:1985,Tremblay:Zanchi:2001}.

\begin{figure}[t]
\centering  \includegraphics*[width=.7\textwidth]{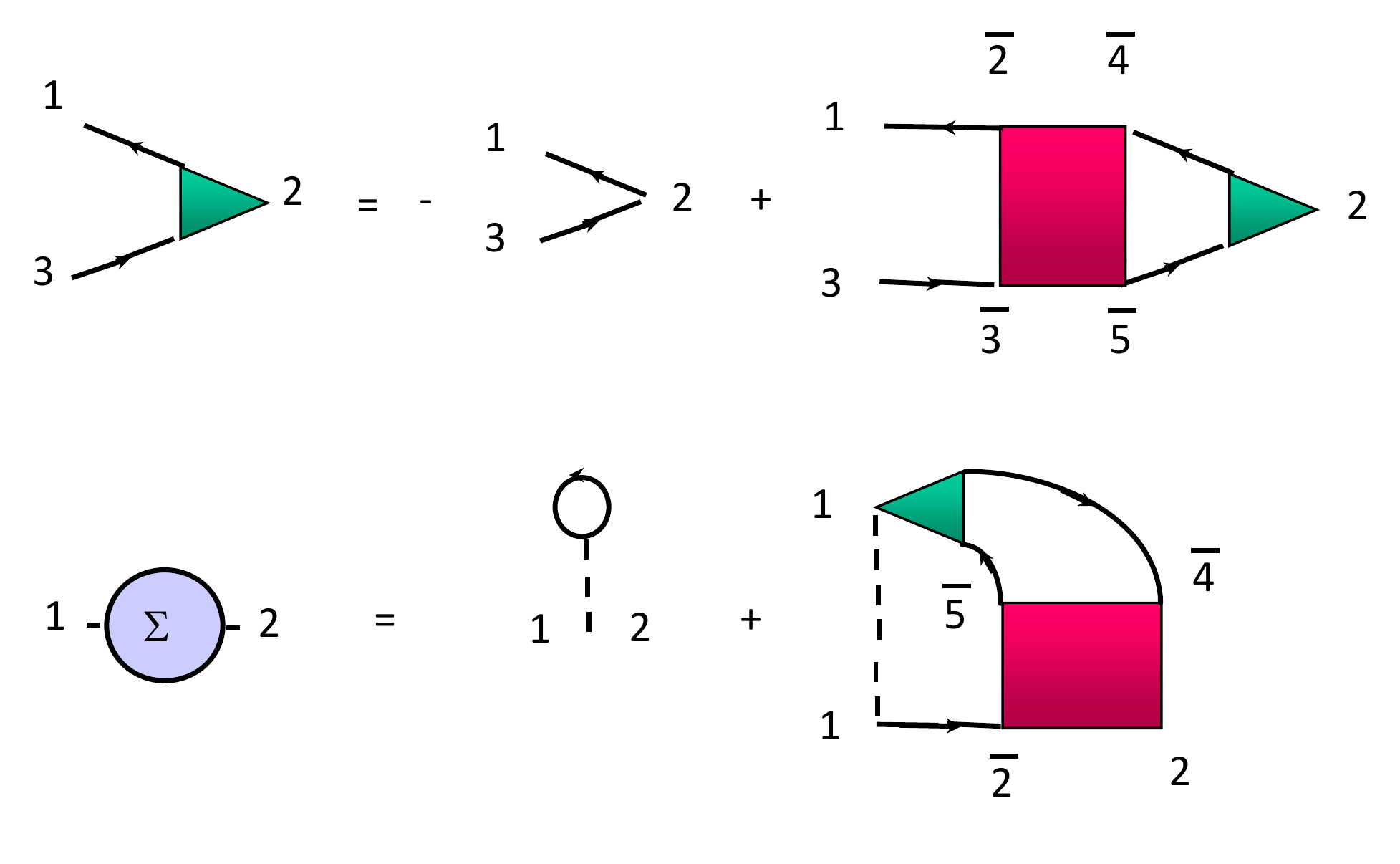}
\caption[Exact diagrammatic representation of response functions and self-energy]{Exact expression for the three point vertex (green triangle) in the
first line and for the self-energy in the second line. Irreducible vertices
are the red boxes and Green's functions solid black lines. The numbers refer
to spin, space and imaginary time coordinates. Symbols with an over-bard are
summed/integrated over. The self-energy is the blue circle and the bare
interaction $U$ the dashed line. }
\label{Tremblay:f_Exact_Self}
\end{figure}

The method to derive the result is justified using the formal derivation
\cite{Tremblay:Allen:2003} that appears in Sect.\ref{Tremblay:Formal}. If you are familiar with diagrams, you can understand physically the
result by looking at Fig. \ref{Tremblay:f_Exact_Self} that shows the exact
diagrammatic expressions for the three-point vertex (green triangle) and
self-energy (blue circle) in terms of Green's functions (solid black lines)
and irreducible vertices (red boxes). The bare interaction $U$ is the dashed
line. One should keep in mind that we are not using perturbation theory
despite the fact that we draw diagrams. Even within an exact approach, the
quantities defined in the figure have well defined meanings. The numbers on
the figure refer to spin, space and imaginary time coordinates. When there
is an over-bar, there is a sum over spin and spatial indices and an integral
over imaginary time.

In TPSC, the irreducible vertices in the first line of Fig. \ref{Tremblay:f_Exact_Self} are local, i.e. completely momentum and frequency
independent. They are given by $U_{sp}$ and $U_{ch}.$ If we set point $3$ to
be the same as point $1,$ then we can obtain directly the TPSC spin and
charge susceptibilities from that first line. In the second line of the
figure, the exact expression for the self-energy is displayed\footnote{In the Hubbard model the Fock term cancels with the same-spin Hartree term}. The first
term on the right-hand side is the Hartree-Fock contribution. In the second
term, one recognizes the bare interaction $U$ at one vertex that excites a
collective mode represented by the green triangle and the two Green's
functions. The other vertex is dressed, as expected. In the electron gas,
the collective mode would be the plasmon. If we replace the irreducible
vertex using $U_{sp}$ and $U_{ch}$ found for the collective modes, we find that
here, both types of modes, spin and charge, contribute to the self-energy \cite{Tremblay:Vilk:1996}.

\begin{figure}[t]
\centering\includegraphics*[width=.7\textwidth]{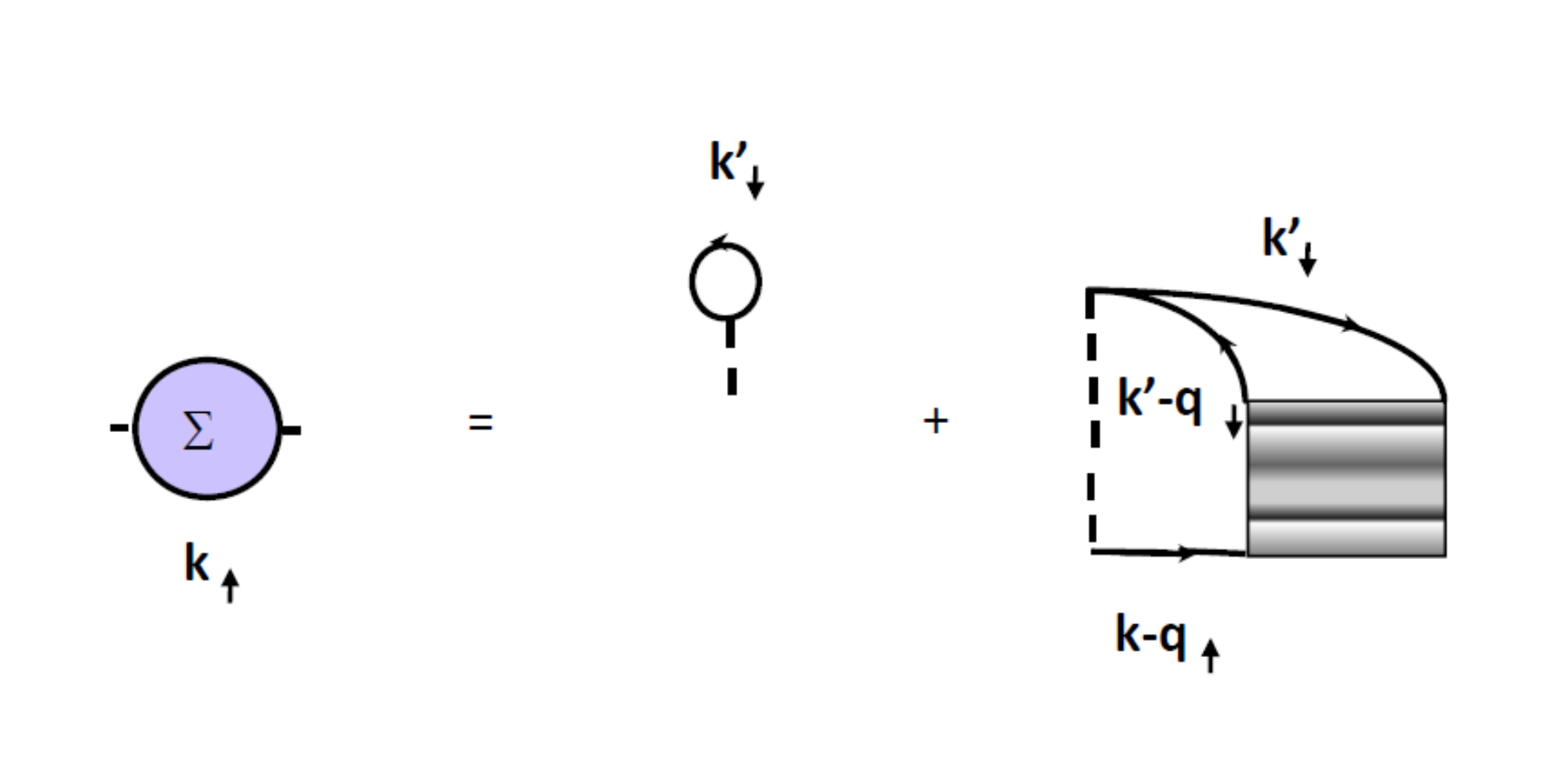}
\caption[Exact self-energy in terms of the Hartree-Fock contribution and of
the fully reducible vertex]{Exact self-energy in terms of the Hartree-Fock contribution and of
the fully reducible vertex $\Gamma $ represented by a textured box.}
\label{Tremblay:f_Exact_Self_full}
\end{figure}

\index{particle-hole channel}
\index{particle-particle channel}
There is, however, an ambiguity in obtaining the self-energy formula \cite{Tremblay:Moukouri:2000}. Within
the assumption that only $U_{sp}$ and $U_{ch}$ enter as irreducible
particle-hole vertices, the self-energy expression in the transverse spin
fluctuation channel is different. What do we mean by that? Consider the
exact formula for the self-energy represented symbolically by the diagram of
Fig.\ \ref{Tremblay:f_Exact_Self_full}. In this figure, the textured box is the fully
reducible vertex $\Gamma \left( q,k-k^{\prime },k+k^{\prime }-q\right) $
that depends in general on three momentum-frequency indices. The
longitudinal version of the self-energy corresponds to expanding the
fully reducible vertex in terms of diagrams that are irreducible in the
longitudinal (parallel spins) channel illustrated in Fig. \ref{Tremblay:f_Exact_Self}%
. This takes good care of the singularity of $\Gamma $ when its first argument $q$
is near $\left( \pi ,\pi \right) .$ The transverse version \cite{Tremblay:Moukouri:2000,Tremblay:Allen:2003} does
the same for the dependence on the second argument $k-k^{\prime }$, which
corresponds to the other (antiparallel spins) particle-hole channel. But the
fully reducible vertex obeys crossing symmetry. In other words,
interchanging two fermions just leads to a minus sign. One then expects that
averaging the two possibilities gives a better approximation for $\Gamma $
since it preserves crossing symmetry in the two particle-hole channels \cite{Tremblay:Moukouri:2000}. By
considering both particle-hole channels only, we neglect the dependence of $%
\Gamma $ on $k+k^{\prime }-q$ because the particle-particle channel is not
singular. The final formula that we obtain is \cite{Tremblay:Moukouri:2000}\
\begin{equation}
\Sigma _{\sigma }^{\left( 2\right) }(k)=Un_{-\sigma }+\frac{U}{8}\frac{T}{N}%
\sum_{q}\left[ 3U_{sp}\chi _{sp}(q)+U_{ch}\chi _{ch}(q)\right] G_{\sigma
}^{\left( 1\right) }(k+q),  \label{Tremblay:Self}
\end{equation}%
where $n_{-\sigma }$ is the average single-spin occupation.
The superscript $\left( 2\right) $ reminds us that we are at the second
level of approximation. $G_{\sigma }^{\left( 1\right) }$ is the same Green's
function as that used to compute the susceptibilities $\chi ^{\left(
1\right) }(q)$. Since the self-energy is constant at that first level of
approximation, this means that $G_{\sigma }^{\left( 1\right) }$ is the
non-interacting Green's function with the chemical potential that gives
the correct filling. That chemical potential $\mu ^{\left( 1\right) }$ is slightly
different from the one that we must use in $\left( G^{\left( 2\right)
}\right) ^{-1}=i\omega _{n}+\mu ^{\left( 2\right) }-\varepsilon _{\mathbf{k}%
}-\Sigma ^{\left( 2\right) }$ to obtain the same density \cite{Tremblay:Kyung:2001}. Estimates of $\mu ^{\left( 1\right) }$
may be found in Ref. \cite{Tremblay:Allen:2003,Tremblay:Kyung:2001}). Further
justifications for the above formula are given below in Sect.\ref{Tremblay:Sec
Internal accuracy}.

\index{spectral weight!QMC,TPSC and FLEX}
\begin{figure}[t]
\centering\includegraphics*[width=1.\textwidth]{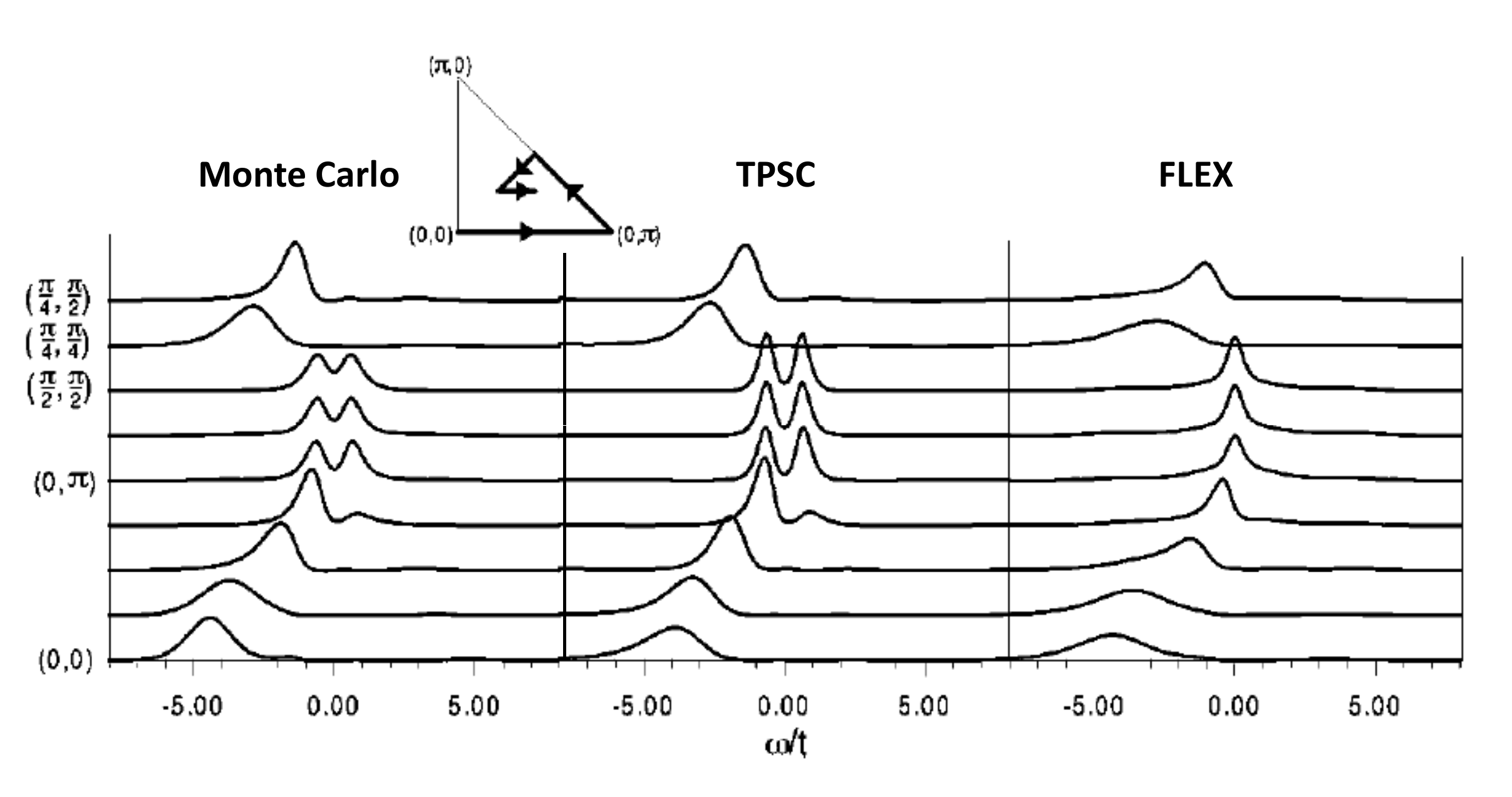}
\caption[Comparisons between QMC, TPSC and FLEX for the pseudogap at half-filling]{Single-particle spectral weight $A(\mathbf{k},\protect\omega )$ for
$U=4$, $\protect\beta =5$, $n=1$, and all independent wave vectors $\mathbf{k%
}$ of an $8\times 8$ lattice. Results obtained from maximum entropy
inversion of Quantum Monte Carlo data on the left panel, from TPSC in the
middle panel and form the FLEX approximation on the right panel. (Relative
error in all cases is about 0.3\%). Figure from Ref.\protect\cite{Tremblay:Moukouri:2000}}
\label{Tremblay:f_QMC_TPSC_FLEX}
\end{figure}

\index{pseudogap}
\index{fluctuation exchange approximation}
But before we come up with more formalism, we check that the above formula
is accurate by comparing in Fig.\ \ref{Tremblay:f_QMC_TPSC_FLEX} the spectral weight
(imaginary part of the Green's function) obtained from Eq.(\ref{Tremblay:Self}) with
that obtained from Quantum Monte Carlo calculations. The latter are exact
within statistical accuracy and can be considered as benchmarks. The meaning
of the curves are detailed in the caption. The comparison is for
half-filling in a regime where the simulations can be done at very low
temperature and where a non-trivial phenomenon, the pseudogap, appears. This
all important phenomenon is discussed further below in subsection \ref{Tremblay:Sec
Pseudogap} and in the first case study, Sect. \ref{Tremblay:Pseudogap in e-doped}. In
the third panel, we show the results of another popular Many-Body Approach,
the FLuctuation Exchange Approximation (FLEX) \cite{Tremblay:Bickers:1989}. It misses \cite{Tremblay:Deisz:1996} the physics of
the pseudogap in the single-particle spectral weight because it uses fully dressed Green's functions and assumes that Migdal's theorem applies, i.e. that the vertex does not need to be renormalized consequently Ref.\cite{Tremblay:Vilk:1997,Tremblay:Monthoux:2003}. The same problem exists in the corresponding version of the GW approximation. \cite{Tremblay:Hedin:1999}

\begin{remark}
The dressing of one vertex in the second line of Fig.\ \ref{Tremblay:f_Exact_Self}
means that we do not assume a Migdal theorem. Migdal's theorem arises in the
case of electron-phonon interactions \cite{Tremblay:Mahan:2000}. There, the small ratio $m/M,$
where $m$ is the electronic mass and $M$ the ionic mass, allows one to show
that the vertex corrections are negligible. This is extremely useful in
formulating the Eliashberg theory of superconductivity.
\end{remark}

\begin{remark}
In Refs. \cite{Tremblay:Vilk:1997,Tremblay:Moukouri:2000} we used the notation $\Sigma ^{\left( 1\right) }$ instead
of $\Sigma ^{\left( 2\right) }.$ The notation of the present paper is the
same as that of Ref. \cite{Tremblay:Allen:2003}
\end{remark}

\subsection{Internal accuracy checks\label{Tremblay:Sec Internal accuracy}}
\index{Two-Particle-Self-Consistent Approach!internal accuracy check}

How can we make sure that TPSC is accurate? We have shown sample
comparisons with benchmark Quantum Monte Carlo calculations, but we can
check the accuracy in other ways. For example, we have already mentioned
that the f-sum rule Eq.(\ref{Tremblay:f-sum q dep}) is exactly satisfied at the first
level of approximation (i.e. with $n_{\mathbf{k}}^{\left( 1\right) }$ on the
right-hand side). Suppose that on the right-hand side of that equation, one
uses $n_{\mathbf{k}}$ obtained from $G^{\left( 2\right) }$ instead of the
Fermi function. One should find that the result does not
change by more than a few percent. This is what happens when agreement with QMC is good.

\index{Luttinger's sum rule}
When we are in the Fermi liquid regime, another way to verify the accuracy
of the approach is to verify if the Fermi surface obtained from $G^{\left(
2\right) }$ satisfies Luttinger's theorem very closely.

Finally, there is a consistency relation between one- and two-particle
quantities ($\Sigma $ and $\left\langle n_{\uparrow }n_{\downarrow
}\right\rangle $). The relation%
\begin{equation}
\frac{T}{N}\sum_\mathbf{k}\sum_n\Sigma(\mathbf{k},i\omega_n)G(\mathbf{k},i\omega_n)e^{-i\omega_n 0^{-}}=\frac{1}{2}\mathrm{Tr}\left( \Sigma G\right) =U\left\langle n_{\uparrow
}n_{\downarrow }\right\rangle
\end{equation}%
should be satisfied exactly for the Hubbard model. This result follows from the definition of self-energy and is derived in Eq.(\ref{Tremblay:SigmaG=un_upn_dwn}) below. In standard many-body
books \cite{Tremblay:Mahan:2000}, it is encountered in the calculation of the free energy
through a coupling-constant integration. In TPSC, it is not difficult
\footnote{Appendix B or Ref. \cite{Tremblay:Vilk:1997}} to show that the following equation%
\begin{equation}
\frac{1}{2}\mathrm{Tr}\left( \Sigma ^{\left( 2\right) }G^{\left( 1\right)
}\right) =U\left\langle n_{\uparrow }n_{\downarrow }\right\rangle
\label{Tremblay:Consistency 1-2}
\end{equation}%
is satisfied exactly with the self-consistent $U\left\langle n_{\uparrow
}n_{\downarrow }\right\rangle $ obtained with the susceptibilities\footnote{FLEX does not satisfy this consistency requirement. See Appendix E of \cite{Tremblay:Vilk:1997}. In fact double-occupancy obtained from $\Sigma G$ can even become negative \cite{Tremblay:Arita:2000}.}. \index{fluctuation exchange approximation} An
internal accuracy check consists in verifying by how much $\frac{1}{2}%
\mathrm{Tr}\left( \Sigma ^{\left( 2\right) }G^{\left( 2\right) }\right) $
differs from $\frac{1}{2}\mathrm{Tr}\left( \Sigma ^{\left( 2\right)
}G^{\left( 1\right) }\right) .$ Again, in regimes where we have agreement
with Quantum Monte Carlo calculations, the difference is only a few percent.

The above relation between $\Sigma $ and $\left\langle n_{\uparrow
}n_{\downarrow }\right\rangle $ gives us another way to justify our
expression for $\Sigma ^{\left( 2\right) }.$ Suppose one starts from Fig.\
\ref{Tremblay:f_Exact_Self} to obtain a self-energy expression that contains
only the longitudinal spin fluctuations and the charge fluctuations, as was
done in the first papers on TPSC \cite{Tremblay:Vilk:1994}. One finds that each of these
separately contributes an amount $U\left\langle n_{\uparrow }n_{\downarrow
}\right\rangle /2$ to the consistency relation Eq.(\ref{Tremblay:Consistency 1-2}).
Similarly, if we work only in the transverse spin channel
\cite{Tremblay:Moukouri:2000,Tremblay:Allen:2003} we find that each of the two transverse spin components
also contributes $U\left\langle n_{\uparrow }n_{\downarrow }\right\rangle /2$
to $\frac{1}{2}\mathrm{Tr}\left( \Sigma ^{\left( 2\right) }G^{\left(
1\right) }\right) .$ Hence, averaging the two expressions also preserves
rotational invariance. In addition, one verifies numerically that the exact
sum rule (Ref. \cite{Tremblay:Vilk:1997} Appendix A)
\begin{equation}
-\int \frac{d\omega ^{\prime }}{\pi }\Sigma _{\sigma }^{\prime \prime
R}\left( \mathbf{k,}\omega ^{\prime }\right) =U^{2}n_{-\sigma }\left(
1-n_{-\sigma }\right)
\end{equation}%
determining the high-frequency behavior is satisfied to a higher degree of
accuracy with the symmetrized self-energy expression Eq. (\ref{Tremblay:Self}).

Eq. (\ref{Tremblay:Self}) for $\Sigma ^{\left( 2\right) }$ is different from so-called
Berk-Schrieffer type expressions \cite{Tremblay:Berk:1966} that do not satisfy \footnote{(See Ref. \cite{Tremblay:Vilk:1997} Appendix E)}
the consistency condition between one- and two-particle properties, $\frac{1%
}{2}\mathrm{Tr}\left( \Sigma G\right) =$ $U\left\langle n_{\uparrow
}n_{\downarrow }\right\rangle .$

\index{conserving approximations}
\index{FLEX|see{fluctuation exchange approximation}}
\index{fluctuation exchange approximation}
\index{thermodynamical consistency}
\begin{remark}
Schemes, such as FLEX, that use on the right-hand side $G^{\left( 2\right) }$
are thermodynamically consistent (Sect. \ref{Tremblay:Conserving}) and might look better.
However, as we just saw, in Fig.\ \ref{Tremblay:f_QMC_TPSC_FLEX}, FLEX misses some important physics. The reason \cite{Tremblay:Vilk:1997} is
that the vertex entering the self-energy in FLEX is not at the same level of
approximation as the Green's functions. Indeed, since the latter contain
self-energies that are strongly momentum and frequency dependent, the
irreducible vertices that can be derived from these self-energies should
also be frequency and momentum dependent, but they are not. In fact they are
the bare vertices. It is as if the quasi-particles had a lifetime while at
the same time interacting with each other with the bare interaction. Using
dressed Green's functions in the susceptibilities with momentum and
frequency independent vertices leads to problems as well. For example, the
conservation law $\chi _{sp,ch}\left( \mathbf{q=0}\text{\textbf{,}}i\omega
_{n}\right) =0$ is violated in that case, as shown in Appendix A of
Ref.\cite{Tremblay:Vilk:1997}. Further criticism of conserving approaches appears in Appendix E of Ref.\cite{Tremblay:Vilk:1997}
and in Ref.\cite{Tremblay:Allen:2003}.
\end{remark}

\subsection{A more formal derivation\label{Tremblay:Formal}}

\index{Two-Particle-Self-Consistent Approach!formal derivation}
Details of a more formal derivation may be found in Ref.\ \cite{Tremblay:Allen:2001}.
For completeness we repeat some of the derivation. The reader more
interested in the physics may skip that section. The first two subsections
present some general formalism that is then used in the following two
subsections to derive TPSC.

\subsubsection{Single-particle properties}

\index{generating functional}
Following functional methods of the Schwinger school\cite{Tremblay:Baym:1962,Tremblay:BaymKadanoff:1962,Tremblay:MartinSchwinger:1959}, we begin with the
generating function with source fields $\phi _{\sigma }$ and field
destruction operators $\psi $ in the grand canonical ensemble
\begin{equation}
\ln Z\left[ \phi \right] =\ln \mathrm{Tr}\left[e^{-\beta \left( \widehat{H}%
-\mu \widehat{N}\right) }\text{T}_{\tau }\left( e^{-\psi _{\overline{\sigma }%
}^{\dagger }\left( \overline{1}\right) \phi _{\overline{\sigma }}\left(
\overline{1},\overline{2}\right) \psi _{\overline{\sigma }}\left( \overline{2%
}\right) }\right) \right]  \label{Tremblay:Generatrice}
\end{equation}
We adopt the convention that $1$ stands for the position and imaginary time
indices $\left( \mathbf{r}_{1},\tau _{1}\right) .$ The over-bar means
summation over every lattice site and integration over imaginary-time from $%
0 $ to $\beta $, and $\overline{\sigma }$ summation over spins. T$_{\tau }$ is the time-ordering operator.

The propagator in the presence of the source field is obtained from
functional differentiation

\begin{equation}
G_{\sigma }\left( 1,2;\{\phi \}\right) =-\left\langle \psi _{\sigma }\left(
1\right) \psi _{\sigma }^{\dagger }\left( 2\right) \right\rangle _{\phi }=-%
\frac{\delta \ln Z\left[ \phi \right] }{\delta \phi _{\sigma }\left(
2,1\right) }.  \label{Tremblay:propagateur}
\end{equation}
From now on, \textit{the time-ordering operator in averages, }$\left\langle
{}\right\rangle $, \textit{is implicit}. Physically relevant correlation
functions are obtained for $\{\phi \}=0$ but it is extremely convenient to
keep finite $\{\phi \}$ in intermediate steps of the calculation.

Using the equation of motion for the field $\psi $ and the definition of the
self-energy, one obtains the Dyson equation in the presence of the source
field \cite{Tremblay:KadanoffLivre:1962}
\begin{equation}
\left( G_{0}^{-1}-\phi \right) G=1+\Sigma G\quad ;\quad
G^{-1}=G_{0}^{-1}-\phi -\Sigma  \label{Tremblay:Dyson}
\end{equation}%
where, from the commutator of the interacting part of the Hubbard
Hamiltonian $H,$ one obtains
\begin{equation}
\Sigma _{\sigma }\left( 1,\overline{1};\{\phi \}\right) G_{\sigma }\left(
\overline{1},2;\{\phi \}\right) =-U\left\langle \psi _{-\sigma }^{\dagger
}\left( 1^{+}\right) \psi _{-\sigma }\left( 1\right) \psi _{\sigma }\left(
1\right) \psi _{\sigma }^{\dagger }\left( 2\right) \right\rangle _{\phi }.
\label{Tremblay:Sigma*G}
\end{equation}%
The imaginary time in $1^{+}$ is infinitesimally larger than in $1$.

\subsubsection{Response functions}

\index{response functions}
Response (four-point) functions for spin and charge excitations can be
obtained from functional derivatives $\left( \delta G/\delta \phi \right) $
of the source-dependent propagator. Following the standard approach and
using matrix notation to abbreviate the summations and integrations we have,

\begin{equation}
GG^{-1}=1  \label{Tremblay:GG-1=1}
\end{equation}
\begin{equation}
\frac{\delta G}{\delta \phi }G^{-1}+G\frac{\delta G^{-1}}{\delta \phi }=0.
\end{equation}
Using the Dyson equation (\ref{Tremblay:Dyson}) $G^{-1}=G_{0}^{-1}-\phi -\Sigma $
this may be rewritten

\begin{equation}
\frac{\delta G}{\delta \phi }=-G\frac{\delta G^{-1}}{\delta \phi }G=G_{%
\symbol{94}}G+G\frac{\delta \Sigma }{\delta \phi }G,  \label{Tremblay:RPA_phi}
\end{equation}
\index{particle-hole channel}
where the symbol $_{\symbol{94}}$ in $G_{\symbol{94}}G$ reminds us that the neighboring labels of
the propagators have to be the same as those of the $\phi $ in the
functional derivative. If perturbation theory converges, we may write the
self-energy as a functional of the propagator$.$ From the chain rule, one
then obtains an integral equation for the response function in the
particle-hole channel that is the analog of the Bethe-Salpeter equation in
the particle-particle channel
\begin{equation}
\frac{\delta G}{\delta \phi }=G_{\symbol{94}}G+G\left[ \frac{\delta \Sigma }{%
\delta G}\frac{\delta G}{\delta \phi }\right] G.  \label{Tremblay:RPA_G}
\end{equation}
The labels of the propagators in the last term are attached to the self
energy, as in Eq.(\ref{Tremblay:RPA_phi}) \footnote{To remind ourselves of this, we may also adopt an additional vertical matrix notation convention and write Eq.(7) as $\frac{\delta G}{\delta \phi }=G_{\symbol{94}}G+G\left[\frac{ \frac{\delta \Sigma }{\delta G}}{\frac{\delta G}{\delta \phi }}\right] G$.}. Vertices appropriate
for spin and charge responses are given, respectively, by
\begin{equation}
U_{sp}=\frac{\delta \Sigma _{\uparrow }}{\delta G_{\downarrow }}-\frac{%
\delta \Sigma _{\uparrow }}{\delta G_{\uparrow }}\quad ;\quad U_{ch}=\frac{%
\delta \Sigma _{\uparrow }}{\delta G_{\downarrow }}+\frac{\delta \Sigma
_{\uparrow }}{\delta G_{\uparrow }}.  \label{Tremblay:dSigma/dG}
\end{equation}

\subsubsection{TPSC First step: two-particle self-consistency for $G^{\left(
1\right) },\Sigma ^{\left( 1\right) },$ $\Gamma _{sp}^{\left( 1\right)
}=U_{sp}$ and $\Gamma _{ch}^{\left( 1\right) }=U_{ch}$ \label{Tremblay:TPSC first
step}}

\index{Pauli principle}
In conserving approximations, the self-energy is obtained from a functional
derivative $\Sigma \left[ G\right] =\delta \Phi \left[ G\right] /\delta G$
of $\Phi $ the Luttinger-Ward functional, which is itself computed from a
set of diagrams. To liberate ourselves from diagrams, we start instead from
the exact expression for the self-energy, Eq.(\ref{Tremblay:Sigma*G}) and notice that
when label $2$ equals $1^{+},$ the right-hand side of this equation is equal
to double-occupancy $\left\langle n_{\uparrow }n_{\downarrow }\right\rangle $%
. Factoring as in Hartree-Fock amounts to assuming no correlations. Instead,
we should insist that $\left\langle n_{\uparrow }n_{\downarrow
}\right\rangle $ be obtained self-consistently. After all, in the
Hubbard model, there are only two local four point functions: $\left\langle
n_{\uparrow }n_{\downarrow }\right\rangle $ and $\left\langle n_{\uparrow
}n_{\uparrow}\right\rangle =\left\langle n_{\downarrow }n_{\downarrow }\right\rangle .$ The
latter is given exactly, through the Pauli principle, by $\left\langle
n_{\uparrow }n_{\uparrow }\right\rangle =\left\langle n_{\downarrow}n_{\downarrow}
\right\rangle =\left\langle n_{\uparrow }\right\rangle =\left\langle
n_{\downarrow }\right\rangle =n/2,$ when the filling $n$ is known$.$ In a
way, $\left\langle n_{\uparrow }n_{\downarrow }\right\rangle $ in the
self-energy equation (\ref{Tremblay:Sigma*G}), can be considered as an initial
condition for the four point function when one of the points, $2$, separates
from all the others which are at $1.$ When that label $2$ does not coincide
with $1$, it becomes more reasonable to factor \textit{\`{a} la}
Hartree-Fock. These physical ideas are implemented by postulating

\begin{subequations}
\begin{equation}
\Sigma _{\sigma }^{\left( 1\right) }\left( 1,\overline{1};\{\phi \}\right)
G_{\sigma }^{\left( 1\right) }\left( \overline{1},2;\{\phi \}\right)
=A_{\{\phi \}}G_{-\sigma }^{\left( 1\right) }\left( 1,1^{+};\{\phi \}\right)
G_{\sigma }^{\left( 1\right) }\left( 1,2;\{\phi \}\right)
\label{Tremblay:Ansatz formal}
\end{equation}%
where $A_{\{\phi \}}$ depends on external field and is chosen such that the
exact result \footnote{See footnote (14) of Ref. \cite{Tremblay:Allen:2003} for a discussion of
the choice of limit $1^+$ vs $1^-$.}
\end{subequations}
\begin{equation}
\Sigma _{\sigma }\left( 1,\overline{1};\{\phi \}\right) G_{\sigma }\left(
\overline{1},1^{+};\{\phi \}\right) =U\left\langle n_{\uparrow }\left(
1\right) n_{\downarrow }\left( 1\right) \right\rangle _{\phi }
\label{Tremblay:SigmaG=un_upn_dwn}
\end{equation}%
is satisfied. It is easy to see that the solution is
\begin{equation}
A_{\{\phi \}}=U\frac{\left\langle n_{\uparrow }\left( 1\right) n_{\downarrow
}\left( 1\right) \right\rangle _{\phi }}{\left\langle n_{\uparrow }\left(
1\right) \right\rangle _{\phi }\left\langle n_{\downarrow }\left( 1\right)
\right\rangle _{\phi }}.
\end{equation}%
Substituting $A_{\{\phi \}}$ back into our \textit{ansatz} Eq.(\ref{Tremblay:ansatz})
we obtain our first approximation for the self-energy by right-multiplying
by $\left( G_{\sigma }^{\left( 1\right) }\right) ^{-1}:$%
\begin{equation}
\Sigma _{\sigma }^{\left( 1\right) }\left( 1,2;\{\phi \}\right) =A_{\{\phi
\}}G_{-\sigma }^{\left( 1\right) }\left( 1,1^{+};\{\phi \}\right) \delta
\left( 1-2\right) .
\end{equation}

We are now ready to obtain irreducible vertices using the prescription of
the previous section, Eq.(\ref{Tremblay:dSigma/dG}), namely through functional
derivatives of $\Sigma $ with respect to $G.$ In the calculation of $U_{sp},$
the functional derivative of $\left\langle n_{\uparrow }n_{\downarrow
}\right\rangle /\left( \left\langle n_{\uparrow }\right\rangle \left\langle
n_{\downarrow }\right\rangle \right) $ drops out, so we are left with \footnote%
{For $n > 1$, all particle occupation numbers must be replaced by hole occupation numbers.}%
\begin{equation}
U_{sp}=\left. \frac{\delta \Sigma _{\uparrow }^{\left( 1\right) }}{\delta
G_{\downarrow }^{\left( 1\right) }}\right\vert _{\{\phi \}=0}-\left. \frac{%
\delta \Sigma _{\uparrow }^{\left( 1\right) }}{\delta G_{\uparrow }^{\left(
1\right) }}\right\vert _{\{\phi \}=0}=A_{\{\phi \}=0}=U\frac{\left\langle
n_{\uparrow }n_{\downarrow }\right\rangle }{\left\langle n_{\uparrow
}\right\rangle \left\langle n_{\downarrow }\right\rangle }.  \label{Tremblay:Usp}
\end{equation}%
The renormalization of this irreducible vertex may be physically understood
as coming from Kanamori-Brueckner screening \cite{Tremblay:Vilk:1997}. This completes
the derivation of the \textit{ansatz} that was missing in our first
derivation in section \ref{Tremblay:Physically-motivated}.

The functional-derivative procedure generates an expression for the charge
vertex $U_{ch}$ which involves the functional derivative of $\left\langle
n_{\uparrow }n_{\downarrow }\right\rangle /\left( \left\langle n_{\uparrow
}\right\rangle \left\langle n_{\downarrow }\right\rangle \right) $ which
contains six point functions that one does not really know how to evaluate.
But, if we again assume that the vertex $U_{ch}$ is a constant, it is simply
determined by the requirement that charge fluctuations also satisfy the
fluctuation-dissipation theorem and the Pauli principle, as in Eq.(\ref{Tremblay:TPSC_charge}).

Note that, in principle, $\Sigma ^{\left( 1\right) }$ also depends on
double-occupancy, but since $\Sigma ^{\left( 1\right) }$ is a constant, it
is absorbed in the definition of the chemical potential and we do not need
to worry about it in this case. That is why the non-interacting irreducible
susceptibility $\chi ^{\left( 1\right) }(q)=\chi _{0}(q)$ appears in the
expressions for the susceptibility, even though it should be evaluated with $%
G^{\left( 1\right) }$ that contains $\Sigma ^{\left( 1\right) }.$ A rough
estimate of the renormalized chemical potential (or equivalently of $\Sigma
^{\left( 1\right) }$), is given in the appendix of Ref.~\cite{Tremblay:Allen:2003}.
One can check that spin and charge conservation are satisfied by our
susceptibilities.

\subsubsection{TPSC Second step: an improved self-energy $\Sigma ^{\left(
2\right) }$}

\index{self-energy!non-perturbative TPSC}
Collective modes are emergent objects that are less influenced by details of the single-particle
properties than the other way around. We thus wish now to obtain an improved
approximation for the self-energy that takes advantage of the fact that we
have found accurate approximations for the low-frequency spin and charge
fluctuations. We begin from the general definition of the self-energy Eq.(%
\ref{Tremblay:Sigma*G}) obtained from Dyson's equation. The right-hand side of that
equation can be obtained either from a functional derivative with respect to
an external field that is diagonal in spin, as in our generating function
Eq.(\ref{Tremblay:Generatrice}), or by a functional derivative of $\left\langle \psi
_{-\sigma }\left( 1\right) \psi _{\sigma }^{\dagger }\left( 2\right)
\right\rangle _{\phi _{t}}$ with respect to a transverse external field $%
\phi _{t}.$

Working first in the longitudinal channel, the right-hand side of the general
definition of the self-energy Eq.(\ref{Tremblay:Sigma*G}) may be written as

\begin{equation}
\Sigma _{\sigma }\left( 1,\overline{1}\right) G_{\sigma }\left( \overline{1}%
,2\right) =-U\left[ \left. \frac{\delta G_{\sigma }\left( 1,2;\{\phi
\}\right) }{\delta \phi _{-\sigma }\left( 1^{+},1\right) }\right\vert
_{\{\phi \}=0}-G_{-\sigma }\left( 1,1^{+}\right) G_{\sigma }\left(
1,2\right) \right] .
\end{equation}%
The last term is the Hartree-Fock contribution. It gives the exact result
for the self-energy in the limit $\omega \rightarrow \infty $.\cite{Tremblay:Vilk:1997} The $\delta G_{\sigma }/\delta \phi _{-\sigma }$ term is thus a
contribution to lower frequencies and it comes from the spin and charge
fluctuations. Right-multiplying the last equation by $G^{-1}$ and replacing
the lower energy part $\delta G_{\sigma }/\delta \phi _{-\sigma }$ by its
general expression in terms of irreducible vertices, Eq.(\ref{Tremblay:RPA_G}) we
find
\begin{eqnarray}
\Sigma _{\sigma }^{\left( 2\right) }\left( 1,2\right) &=&UG_{-\sigma
}^{\left( 1\right) }\left( 1,1^{+}\right) \delta \left( 1-2\right)
\label{Tremblay:SigmaLongExact} \\
&&-UG_{\sigma }^{\left( 1\right) }\left( 1,\overline{3}\right) \left[ \left.
\frac{\delta \Sigma _{\sigma }^{\left( 1\right) }\left( \overline{3}%
,2;\{\phi \}\right) }{\delta G_{\overline{\sigma }}^{\left( 1\right) }\left(
\overline{4},\overline{5};\{\phi \}\right) }\right\vert _{\{\phi \}=0}\left.
\times\frac{\delta G_{\overline{\sigma }}^{\left( 1\right) }\left( \overline{4},%
\overline{5};\{\phi \}\right) }{\delta \phi _{-\sigma }\left( 1^{+},1\right)
}\right\vert _{\{\phi \}=0}\right] .  \notag
\end{eqnarray}%
Every quantity appearing on the right-hand side of that equation has been
taken from the TPSC results. This means in particular that the irreducible
vertices $\delta \Sigma _{\sigma }^{\left( 1\right) }/\delta G_{\sigma
^{\prime }}^{\left( 1\right) }$ are at the same level of approximation as
the Green functions $G_{\sigma }^{\left( 1\right) }$ and self-energies $%
\Sigma _{\sigma }^{\left( 1\right) }.$ In approaches that assume that
Migdal's theorem applies to spin and charge fluctuations, one often sees
renormalized Green functions $G^{\left( 2\right) }$ appearing on the
right-hand side along with unrenormalized vertices, $\delta \Sigma _{\sigma
}/\delta G_{\sigma ^{\prime }}\rightarrow U.$ In terms of $U_{sp}$ and $%
U_{ch}$ in Fourier space, the above formula\cite{Tremblay:Vilk:1996} reads,

\begin{equation}
\Sigma _{\sigma }^{\left( 2\right) }(k)_{long}=Un_{-\sigma }+\frac{U}{4}%
\frac{T}{N}\sum_{q}\left[ U_{sp}\chi _{sp}^{\left( 1\right) }(q)+U_{ch}\chi
_{ch}^{\left( 1\right) }(q)\right] G_{\sigma }^{\left( 1\right) }(k+q).
\label{Tremblay:Self-longitudinal}
\end{equation}

The approach to obtain a self-energy formula that takes into account
both longitudinal and transverse fluctuations is detailed in Ref.~\cite{Tremblay:Allen:2003}. Crossing symmetry, rotational
symmetry and sum rules and comparisons with QMC dictate the final formula for the improved
self-energy $\Sigma ^{\left( 2\right) }$ as we have explained in Sect.(\ref{Tremblay:Sec-Self}).

\subsection{Pseudogap in the renormalized classical regime\label{Tremblay:Sec
Pseudogap}}

\index{renormalized classical regime}
\index{pseudogap}
\begin{figure}[t]
\centering\includegraphics*[width=.7\textwidth]{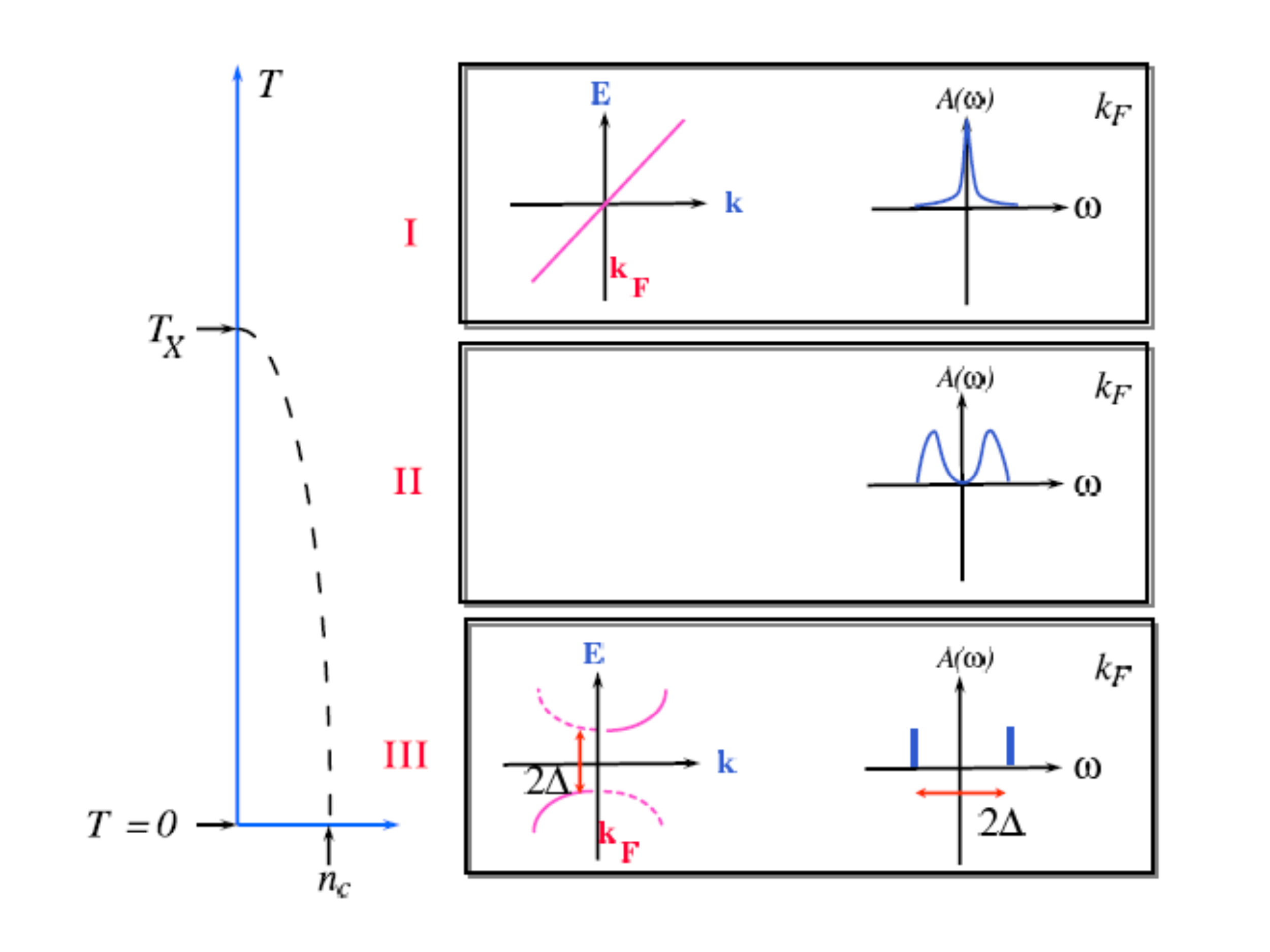}
\caption[Cartoon explanation of the pseudogap due to precursors of long-range order]{Cartoon explanation of the pseudogap due to precursors of long-range order. When the
antiferromagnetic correlation length $\protect\xi$ becomes larger than the
thermal de Broglie wavelength, there appears precursors of the $T=0$
Bogoliubov quasi-particles for the long-range ordered antiferromagnet. This
can occur only in the renormalized classical regime, below the dashed
line on the left of the figure.}
\label{Tremblay:f_88}
\end{figure}

\index{fluctuation exchange approximation}
When we compared TPSC with Quantum Monte Carlo simulations and with FLEX in
Fig.\ \ref{Tremblay:f_QMC_TPSC_FLEX} above, perhaps you noticed that at the Fermi
surface, the frequency dependent spectral weight has two peaks instead of
one. In addition, at zero frequency, it has a minimum instead of a maximum.
That is called a pseudogap. A cartoon explanation \cite{Tremblay:LTP:2006}
of this pseudogap is given in Fig.~\ref{Tremblay:f_88}. At high temperature we start
from a Fermi liquid, as illustrated in panel I. Now, suppose the ground
state has long-range antiferromagnetic order as in panel III, in other words
at a filling between half-filling and $n_{c}$. In the mean-field
approximation we have a gap and the Bogoliubov transformation from fermion
creation-annihilation operators to quasi-particles has weight at both
positive and negative energies. In two dimensions, because of the
Mermin-Wagner theorem, as soon as we raise the temperature above zero,
long-range order disappears, but the antiferromagnetic correlation length $%
\xi $ remains large so we obtain the pseudogap illustrated in panel II. As
we will explain analytically below, the pseudogap survives as long as $\xi $
is much larger than the thermal de Broglie wave length $\xi _{th}\equiv
v_{F}/(\pi T)$ in our usual units. At the crossover temperature $T_{X}$, the
relative size of $\xi $ and $\xi _{th}$ changes and we recover the Fermi
liquid.

\index{renormalized classical regime}
We now proceed to sketch analytically where these results come from
starting from finite $T$. Details and more complete formulae may be found in
Refs.~\cite{Tremblay:Vilk:1994,Tremblay:Vilk:1996,Tremblay:Vilk:1997,Tremblay:Vilk:1995}\footnote{Note also the
following study from zero temperature \cite{Tremblay:borejsza:2004}}. We
begin from the TPSC expression (\ref{Tremblay:Self}) for the self-energy. Normally
one has to do the sum over bosonic Matsubara frequencies first, but the zero
Matsubara frequency contribution has the correct asymptotic behavior in
fermionic frequencies $i\omega_{n}$ so that, as in Sect.\ref{Tremblay:Mermin Wagner}, one can
once more isolate on the
right-hand side the contribution from the zero Matsubara frequency. In the
renormalized classical regime then, we have
\begin{equation}
\Sigma (\mathbf{k}_{F},i\omega_{n})\propto T\int q^{d-1}dq\frac{1}{q^{2}+\xi ^{-2}%
}\frac{1}{i\omega_{n}-\varepsilon _{\mathbf{k}_{F}+\mathbf{Q+q}}}
\label{Tremblay:RC-contribution-sigma}
\end{equation}%
where $\mathbf{Q}$ is the wave vector of the instability.
\footnote{This formula is similar to one that appeared in Ref.\cite{Tremblay:Lee:1973}} This integral can
be done analytically in two dimensions \cite{Tremblay:Vilk:1997,Tremblay:VilkShadow:1997}. But it is
more useful to analyze limiting cases \cite{Tremblay:Vilk:1996}. Expanding around the points known as hot spots
where $\varepsilon _{\mathbf{k}_{F}+\mathbf{Q}}=0$, we find after analytical
continuation that the imaginary part of the retarded self-energy at zero
frequency takes the form
\begin{align}
\Sigma^{\prime \prime R}(\mathbf{k}_{F},0)& \propto -\pi T\int
d^{d-1}q_{\perp }dq_{||}\frac{1}{q_{\perp }^{2}+q_{||}^{2}+\xi ^{-2}}\delta
(v_{F}^{\prime }q_{||}) \\
& \propto \frac{\pi T}{v_{F}^{\prime }}\xi ^{3-d}.  \label{Tremblay:ImSigma2}
\end{align}%
In the last line, we just used dimensional analysis to do the integral.

The importance of dimension comes out clearly \cite{Tremblay:Vilk:1996}. In $d=4$, $\Sigma^{\prime
\prime R}(\mathbf{k}_{F},0)$ vanishes as temperature decreases, $d=3$ is the
marginal dimension and in $d=2$ we have that $\Sigma^{\prime \prime R}(%
\mathbf{k}_{F},0)\propto \xi /\xi _{th}$ that diverges at zero temperature.
In a Fermi liquid the quantity $\Sigma^{\prime \prime R}(\mathbf{k}_{F},0)$
vanishes at zero temperature, hence in three or four dimensions one recovers
the Fermi liquid (or close to one in $d=3$). But in two dimensions, a
diverging $\Sigma^{\prime \prime R}(\mathbf{k}_{F},0)$ corresponds to a
vanishingly small $A(\mathbf{k}_{F},\omega =0)$ as we can see from%
\begin{equation}
A(\mathbf{k},\omega )=\frac{-2\Sigma^{\prime \prime R}(\mathbf{k}_{F},\omega
)}{(\omega -\varepsilon _{\mathbf{k}}-\Sigma^{\prime R}(\mathbf{k}%
_{F},\omega ))^{2}+\Sigma^{\prime \prime R}(\mathbf{k}_{F},\omega )^{2}}.
\label{Tremblay:Spectral-weight_1}
\end{equation}%
Fig. 31 of Ref.\cite{Tremblay:LTP:2006} illustrates graphically the relationship
between the location of the pseudogap and large scattering rates at the
Fermi surface. At stronger $U$ the scattering rate is large over a broader
region, leading to a depletion of $A(\mathbf{k,}\omega )$ over a broader
range of $\mathbf{k}$ values.

\index{renormalized classical regime}
\begin{remark}
Note that the condition $\xi /\xi _{th}\gg 1$, necessary to obtain a large
scattering rate, is in general harder to satisfy than the condition that
corresponds to being in the renormalized classical regime. Indeed, $\xi /\xi
_{th}\gg 1$ corresponds $T/v_{F}\gg \xi ^{-1}$ while the condition $\omega
_{sp}\ll T$ for the renormalized classical regime corresponds to $T\gg \xi
^{-2},$ with appropriate scale factors, because $\omega _{sp}$ scales as $%
\xi ^{-2}$ as we saw in Eq.\ (\ref{Tremblay:Chi_sp_low_frequency}) and below.
\end{remark}

\index{fluctuation exchange approximation}
To understand the splitting into two peaks seen in Figs.~\ref{Tremblay:f_QMC_TPSC_FLEX}
and \ref{Tremblay:f_88} consider the singular renormalized contribution coming from
the spin fluctuations in Eq.~(\ref{Tremblay:RC-contribution-sigma}) at frequencies $%
\omega \gg v_{F}\xi ^{-1}.$ Taking into account that contributions to the
integral come mostly from a region $q\leq \xi ^{-1}$, one finds%
\begin{align}
\Sigma ^{\prime R}(\mathbf{k}_{F},\omega )& =\left( T\int q^{d-1}dq\frac{1}{%
q^{2}+\xi ^{-2}}\right) \frac{1}{ik_{n}-\varepsilon _{\mathbf{k}_{F}+\mathbf{%
Q}}}  \notag \\
& \equiv \frac{\Delta ^{2}}{\omega -\varepsilon _{\mathbf{k}_{F}+\mathbf{Q}}}
\label{Tremblay:self-singular}
\end{align}%
which, when substituted in the expression for the spectral weight (\ref{Tremblay:Spectral-weight_1}) leads to large contributions when
\begin{equation}
\omega -\varepsilon _{\mathbf{k}}-\frac{\Delta ^{2}}{\omega -\varepsilon _{%
\mathbf{k}_{F}+\mathbf{Q}}}=0
\end{equation}%
or, equivalently,%
\begin{equation}
\omega =\frac{(\varepsilon _{\mathbf{k}}+\varepsilon _{\mathbf{k}_{F}+%
\mathbf{Q}})\pm \sqrt{(\varepsilon _{\mathbf{k}}-\varepsilon _{\mathbf{k}%
_{F}+\mathbf{Q}})^{2}+4\Delta ^{2}}}{2},
\end{equation}%
which, at $\omega =0$, corresponds to the position of the hot spots\footnote{For comparisons with paramagnon theory see \cite{Tremblay:Saikawa:2001}.}. At
finite frequencies, this turns into the dispersion relation for the
antiferromagnet \cite{Tremblay:Zhang:1989}.

It is important to understand that analogous arguments hold for any
fluctuation that becomes soft because of the Mermin-Wagner theorem,\cite{Tremblay:Vilk:1997,Tremblay:Davoudi:2008}
including superconducting ones \cite{Tremblay:Vilk:1997,Tremblay:Allen:2001,Tremblay:Kyung:2001}. The wave vector $\mathbf{Q}$ would be
different in each case.

To understand better when Fermi liquid theory is valid and when it is
replaced by the pseudogap instead, it is useful to perform the calculations
that lead to $\Sigma ^{\prime \prime R}(\mathbf{k}_{F},0)\propto \xi /\xi
_{th}$ in the real frequency formalism. The details may be found in Appendix
D of Ref. \cite{Tremblay:Vilk:1997}.

\section{Case studies \label{Tremblay:Case}}

In this short pedagogical review it is impossible to cover all topics in
depth. This section will nevertheless expand a bit on two important
contributions of TPSC to problems of current interest, namely the pseudogap
of cuprate superconductors and superconductivity induced by
antiferromagnetic fluctuations.

\subsection{Pseudogap in electron-doped cuprates\label{Tremblay:Pseudogap in e-doped}}

\index{cuprates}
\index{pseudogap}
High-temperature superconductors are made of layers of CuO$_{2}$ planes. The
rest of the structure is commonly considered as providing either electron or
hole doping of these planes depending on chemistry. At half-filling, or
zero-doping, the ground state is an antiferromagnet. As one dopes the
planes, one reaches a doping, so-called optimal doping, where the
superconducting transition temperature $T_{c}$ is maximum. Let us start from
optimal hole or electron doping and decrease doping towards half-filling.
That is the underdoped regime. In that regime, one observes a curious
phenomenon, the pseudogap. What this means is that as temperature decreases,
physical quantities behave as if the density of states near the Fermi level
were decreasing. Finding an explanation for this phenomenon has been one of
the major challenges of the field \cite{Tremblay:Timusk:1999,Tremblay:Norman:2005}.

To make progress, we need a microscopic model for high-temperature
superconductors. Band structure calculations \cite{Tremblay:Andersen:1995,Tremblay:Andersen:2001} reveal that a
single band crosses the Fermi level. Hence, it is a common assumption that
these materials can be modeled by the one-band Hubbard model. Whether this is an oversimplification
is still a subject of controversy \cite{Tremblay:peets:2009,Tremblay:Liebsch:2010,Tremblay:PhillipsComment:2010,Tremblay:Varma:2009,Tremblay:Macridin:2005,Tremblay:Hanke:2010}.
Indeed, spectroscopic studies \cite{Tremblay:ChenX:1991,Tremblay:peets:2009} show that hole doping
occurs on the oxygen atoms. The resulting hole behaves as a copper excitation because
of Zhang-Rice \cite{Tremblay:Zhang:1988a} singlet formation. In addition, the phase
diagram \cite{Tremblay:Senechal:2005,Tremblay:maier_d:2005,Tremblay:Aichhorn:2006,Tremblay:Aichhorn:2007,Tremblay:Haule:2007,Tremblay:kancharla:2008} and many
properties of the hole-doped cuprates can be described by the one-band
Hubbard model. Typically, the band parameters that are
used are: nearest-neighbor hopping $t=350\;$ to $400\;$meV and
next-nearest-neighbor hopping $t^{\prime }=-0.15$ to $-0.3t$ depending on
the compound \cite{Tremblay:Andersen:1995,Tremblay:Andersen:2001}. A third-nearest-neighbor hopping $t^{\prime \prime
}=-0.5t^{\prime }$ is sometimes added to fit finer details of the band
structure \cite{Tremblay:Andersen:2001}. The second-neighbor hopping breaks particle-hole
symmetry at the band structure level.

\index{ARPES}
In electron-doped cuprates, the doping occurs on the copper,
hence there is little doubt that the single-band Hubbard model is even a
better starting point in this case. Band parameters \cite{Tremblay:Massida:1989} are similar to those of hole-doped cuprates.
It is sometimes claimed that there is a pseudogap only in the hole-doped cuprates. The origin of
the pseudogap is indeed probably different in the
hole-doped cuprates. But even though the standard signature of a pseudogap is
absent in nuclear magnetic resonance \cite{Tremblay:Zheng:2003} (NMR) there is definitely a
pseudogap in the electron-doped case as well \cite{Tremblay:Armitage:2010},
as can be seen in optical conductivity \cite{Tremblay:Onose:2001} and in Angle
Resolved Photoemission Spectroscopy (ARPES) \cite{Tremblay:Armitage:2002}. As we show in
the rest of this section, in electron-doped cuprates strong evidence for the
origin of the pseudogap is provided by detailed comparisons of TPSC with
ARPES as well as by verification with neutron scattering \cite{Tremblay:Motoyama:2007}
that the TPSC condition for a pseudogap, namely $\xi >\xi _{th},$ is
satisfied. The latter length makes sense from weak to intermediate coupling
when quasi-particles exist above the pseudogap temperature. In strong
coupling, i.e. for values of $U$ larger than that necessary for the Mott
transition, there is evidence that there is another mechanism for the
formation of a pseudogap. This is discussed at length in Refs. \cite{Tremblay:Senechal:2004,Tremblay:Hankevych:2006} \footnote{See also conclusion of Ref.\cite{Tremblay:LTP:2006}.}. The recent discovery \cite{Tremblay:Sordi:2010} that at sufficiently
large $U$ there is a first order transition in the paramagnetic state between
two kinds of metals, one of which is highly anomalous, gives a sharper
meaning to what is meant by strong-coupling pseudogap.

\index{superconductivity!and pressure dependence}
Let us come back to modeling of electron-doped cuprates. Evidence that these
are less strongly coupled than their hole-doped counterparts comes from the
fact that a) The value of the optical gap at half-filling, $\sim 1.5$\ eV,
is smaller than for hole doping, $\sim 2.0$\ eV \cite{Tremblay:Tokura:1990}. b) In a simple Thomas-Fermi
picture, the screened interaction scales like $\partial \mu /\partial n.$
Quantum cluster calculations \cite{Tremblay:Senechal:2004} show that $\partial \mu
/\partial n$ is smaller on the electron-doped side, hence $U$ should be
smaller. c) Mechanisms based on the exchange of antiferromagnetic
fluctuations with $U/t$ at weak to intermediate coupling \cite{Tremblay:Bickers_dwave:1989,Tremblay:Kyung:2003}
predict that the superconducting $T_{c}$ increases with $U/t$. Hence $T_{c}$
should decrease with increasing pressure in the simplest model where
pressure increases hopping $t$ while leaving $U$ essentially unchanged. The
opposite behavior, expected at strong coupling where $J=4t^{2}/U$ is
relevant \cite{Tremblay:kancharla:2008,Tremblay:Kotliar:1988}, is observed in the hole-doped cuprates. d)
Finally and most importantly, we have shown detailed agreement between TPSC
calculations \cite{Tremblay:Kyung:2004,Tremblay:Hankevych:2006,Tremblay:LTP:2006} and measurements such as ARPES
\cite{Tremblay:Armitage:2002,Tremblay:Matsui:2005}, optical conductivity \cite{Tremblay:Onose:2001} and neutron \cite{Tremblay:Motoyama:2007}
scattering.
\begin{figure}[tbp]
\centering  \includegraphics*[width=.95\textwidth]{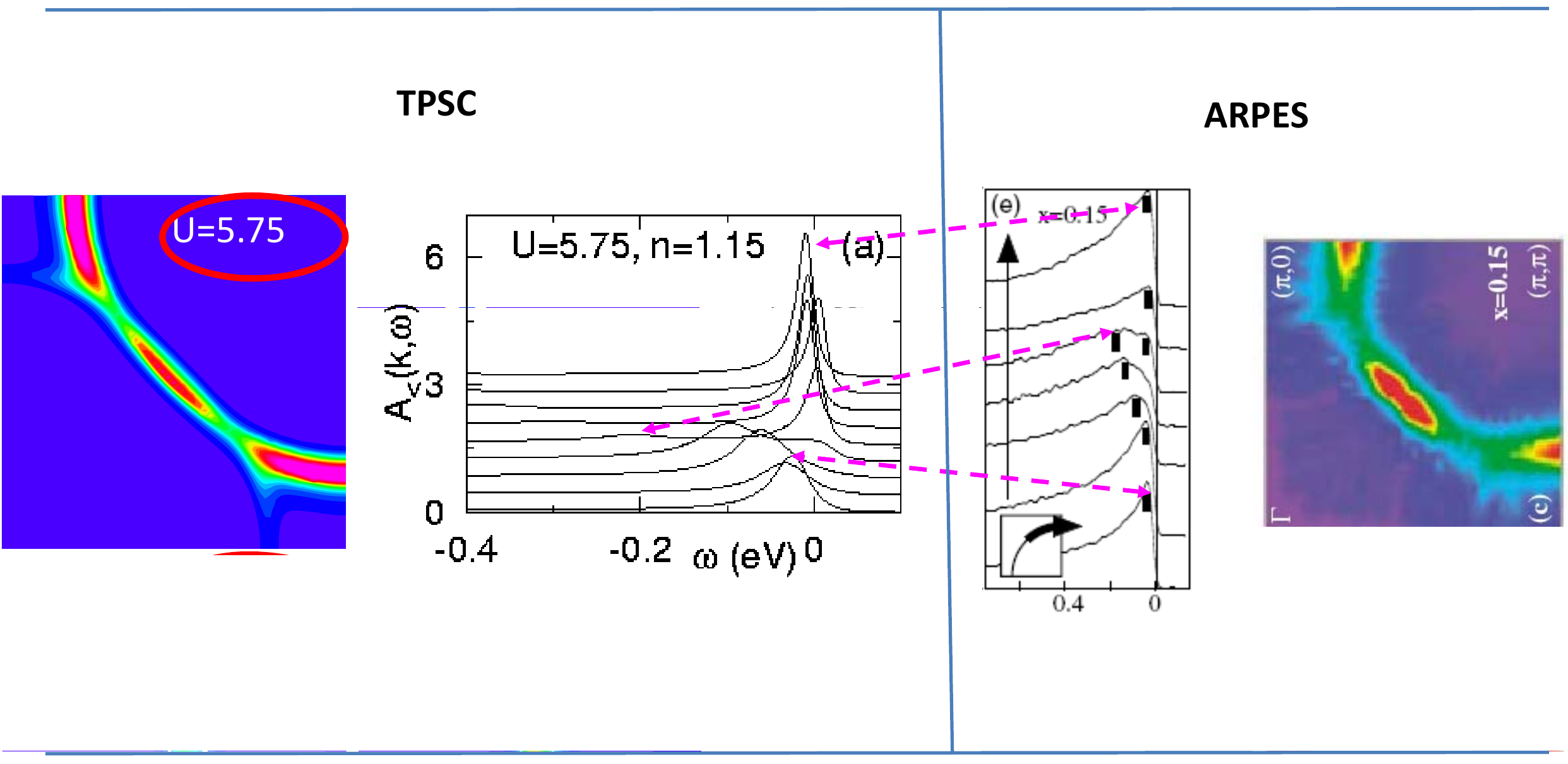}
\caption[Momentum distribution curves and energy distribution curves, TPSC vs ARPES experiment]{On the left, results of TPSC calculations \cite{Tremblay:Kyung:2004,Tremblay:LTP:2006}
at optimal doping, $x=0.15,$ corresponding to filling $1.15,$
for $t=350$ meV, $t^{\prime }=-0.175t,$ $t\textquotedblright =0.05t,$ $%
U=5.75t,$ $T=1/20.$ The left-most panel is the magnitude of the spectral
weight times a Fermi function, $A\left( \mathbf{k},\protect\omega \right)
f\left( \protect\omega \right) $ at $\protect\omega =0,$ so-called
momentum-distribution curve (MDC). Red (dark black) indicates larger value
and purple (light grey) smaller value. The next panel is $A\left( \mathbf{k},%
\protect\omega \right) f\left( \protect\omega \right) $ for a set of fixed $%
\mathbf{k}$ values along the Fermi surface. These are so-called
energy-dispersion curves (EDC). The two panels to the right are the
corresponding experimental results \cite{Tremblay:Armitage:2002} for Nd$_{2-x}$Ce$_{x}$CuO$%
_{4}.$ Dotted arrows show the correspondence between TPSC and experiment. }
\label{Tremblay:f_ARPES_TPSC}
\end{figure}

To illustrate the last point, consider Fig.\ \ref{Tremblay:f_ARPES_TPSC} that
compares TPSC calculations with experimental results for ARPES. Apart from a
tail in the experimental results. the agreement is striking. \footnote{Such tails
tend to disappear in more recent laser ARPES measurements on hole-doped compounds \cite{Tremblay:Koralek:2006}.}. In
particular, if there were no interaction, the Fermi surface would be a line (red)
on the momentum distribution curve (MDC). Instead, it seems to
disappear at symmetrical points displaced from $\left( \pi /2,\pi /2\right)
. $ These points, so-called hot spots, are linked by the wave vector $\left(
\pi ,\pi \right) $ to other points on the Fermi surface. This is where the
antiferromagnetic gap would open first if there were long-range order. The
pull back of the weight from $\omega =0$ at the hot spots is close to the
experimental value: $100$ meV for the $15\%$ doping shown, and $300$ meV for
$10\%$ doping (not shown). More detailed ARPES spectra and comparisons with
experiment are shown in Ref. \cite{Tremblay:LTP:2006}. The value of the temperature $%
T^{\ast }$ at which the pseudogap appears \cite{Tremblay:Kyung:2004} is also close to
that observed in optical spectroscopy \cite{Tremblay:Onose:2001}. In addition, the size of
the pseudogap is about ten times $T^{\ast }$ in the calculation as well as
in the experiments. For optical spectroscopy, vertex corrections (see
Sect. \ref{Tremblay:Vertex}) have to be added to be more quantitative. Experimentally, the value
of $T^{\ast }$ is about twice the antiferromagnetic transition temperature
up to $x=0.13$. That can be obtained \cite{Tremblay:Kyung:2004} by taking $t_{z}=0.03t$
for hopping in the third direction. Recall that in strictly two dimensions,
there is no long-range order. Antiferromagnetism appears on a much larger
range of dopings for electron-doped than for hole-doped cuprates.

These TPSC calculations have predicted the value of the pseudogap
temperature at $x=0.13$ before it was observed experimentally \cite{Tremblay:Matsui:2005} by
a group unaware of the theoretical prediction. In addition, the prediction
that $\xi $ should scale like $\xi _{th}$ at the pseudogap temperature has
been verified in neutron scattering experiments \cite{Tremblay:Motoyama:2007} in the range $%
x=0.04$ to $x=0.15$. At that doping, which corresponds to optimal doping, $%
T^{\ast }$ becomes of the order of $100$ K, more than four times lower than
at $x=0.04.$ The antiferromagnetic correlation length $\xi $ beyond optimal
doping begins to decrease and violate the scaling of $\xi $ with $\xi _{th}.$
In that doping range, $T^{\ast }$ and the superconducting transition
temperature are close. Hence it is likely that there is interference between
the two phenomena \cite{Tremblay:Sedeki:2009}, an effect that has not yet been taken into
account in TPSC.

An important prediction that one should verify is that inelastic neutron
scattering will find over-damped spin fluctuations in the pseudogap regime
and that the characteristic spin fluctuation energy will be smaller than $%
k_{B}T$ whenever a pseudogap is present. Equality should occur above $%
T^{\ast }$.

Finally, note that the agreement found in Fig.\ \ref{Tremblay:f_ARPES_TPSC} between
ARPES and TPSC is for $U\sim 6t.$ At smaller values of $U$ the
antiferromagnetic correlations are not strong enough to produce a pseudogap
in that temperature range. For larger $U,$ the weight near $\left( \pi /2,\pi
/2\right) $ disappears, in disagreement with experiments. The same value of $%
U$ is found for the same reasons in strong coupling calculations with
Cluster Perturbation Theory (CPT) \cite{Tremblay:Senechal:2004} and with slave boson methods
\cite{Tremblay:Yuan:2005}. Recent first principle calculations \cite{Tremblay:Weber:2010} find essentially
the same value of $U.$ In that approach, the value of $U$ is fixed, whereas
in TPSC it was necessary to increase $U$ by about $10\%$ moving towards
half-filling to get the best agreement with experiment. In any case, it is
quite satisfying that weak and strong coupling methods agree on the value of
$U$ for electron-doped cuprates. This value of $U$ is very near the critical
value for the Mott transition at half-filling \cite{Tremblay:park:2008}. Hence,
antiferromagnetic fluctuations at finite doping can be very well described
by Slater-like physics (nesting) in electron-doped cuprates.

For recent calculations including the effect of the third dimension on the pseudogap see \cite{Tremblay:Sedrakyan:2010}.
Finally, note that the analog of the above mechanism for the pseudogap has also been seen
in two-dimensional charge-density wave dichalcogenides \cite{Tremblay:Borisenko:2008}.

\subsection{d-wave superconductivity\label{Tremblay:d-wave}}

\index{superconductivity!d-wave by antiferromagnetic fluctuations}
In the BCS theory of superconductivity, pairs of electrons form because of
an effective attraction mediated by phonons. The pairs then condense in a
coherent state. The suggestion that superconductivity could arise from
purely repulsive forces goes back to Kohn and Luttinger who showed that for
pairs of sufficiently high angular momentum, the \textit{screened} Coulomb
interaction in an electron gas could be attractive \cite{Tremblay:kohn:1965}. Just before the
discovery of high-temperature superconductors, an extension of that idea was
proposed \cite{Tremblay:Beal-Monod:1986,Tremblay:Scalapino:1986,Tremblay:Miyake:1986}. The suggestion
was that antiferromagnetic fluctuations present in the Hubbard model could
replace the phonons in BCS theory and lead to $d-$wave superconductivity.
This is difficult to prove beyond any doubt since superconductivity in this case does not arise at the
mean-field level. Mean-field on the Hubbard model gives antiferromagnetism
near half-filling but not superconductivity. In the high-temperature
superconductors, the situation is made even more difficult because of Mott
Physics. Nevertheless, the question is well posed and, as we just saw, Mott
physics might be less important in the electron-doped superconductors.

To investigate how the pairing susceptibility is influenced by
antiferromagnetic fluctuations in TPSC we proceed as follows \cite{Tremblay:Kyung:2003,Tremblay:Hassan:2008}.
The reader who did not go through the formal section \ref{Tremblay:Formal}
may skip the next paragraph without loss of continuity to read the physical
results below. The few equations that appear below give details that are
missing in the literature.

\begin{description}
\item[Some details of the derivation:] We work in Nambu space and add an
off-diagonal source field $\theta $ and $\theta ^{\ast }$ in the generating
function Eq.(\ref{Tremblay:Generatrice}). The transverse spin fluctuations are
included by working with four by four matrices. The pair susceptibility in
the normal state can be obtained from the second functional derivative of
the generating function with respect to the off-diagonal source field,
evaluated at zero source field. \cite{Tremblay:Kyung:2003} In more detail, one
proceeds as in the formal derivation in Sect.\ \ref{Tremblay:Formal}. The expressions
for the spin and charge susceptibilities are not modified. Once the
two-particle quantities have been found as above, the second step of the
approach \cite{Tremblay:Vilk:1997,Tremblay:Moukouri:2000} consists in improving the
approximation for the single-particle self-energy by starting from the
\textit{exact} expression where the high-frequency Hartree-Fock behavior is
explicitly subtracted
\begin{eqnarray}
\mathbf{\Sigma }\left( 1,2\right)  &=&-U\left(
\begin{tabular}{ll}
$1$ & $0$ \\
$0$ & $0$%
\end{tabular}%
 \right) \frac{\delta \mathbf{G}\left( 1,\overline{3}\right) }{\delta
\phi _{\downarrow }\left( 1^{+},1\right) }\mathbf{G}^{-1}\left( \overline{3}%
,2\right)   \label{Tremblay:self-exact} \\
&&+U\left(
\begin{tabular}{ll}
$0$ & $0$ \\
$0$ & $1$%
\end{tabular}%
 \right) \frac{\delta \mathbf{G}\left( 1,\overline{3}\right) }{\delta
\phi _{\uparrow }\left( 1^{+},1\right) }\mathbf{G}^{-1}\left( \overline{3}%
,2\right) .  \notag
\end{eqnarray}%
The bold face objects are matrices in Nambu space. To be able to express the
right-hand side of the above equation in terms of irreducible vertices,
susceptibilities and powers of $G,$ one differentiates $\mathbf{GG}^{-1}=%
\mathbf{I}$ to obtain $\left( \delta \mathbf{G/}\delta \phi \right) \mathbf{G%
}^{-1}=-\mathbf{G}\left( \delta \mathbf{G}^{-1}\mathbf{/}\delta \phi \right)
.$ With the help of Dyson's equation on the right-hand side of the last
equation as well as the chain rule, one finds an expression where one can
replace every term by their value at the first step, namely $U_{sp}$ and $%
U_{ch}$ for the irreducible low-frequency vertices as well as $G_{\sigma
}^{\left( 1\right) }(k+q)$ and $\chi _{sp}(q),\chi _{ch}(q).$ For the
diagonal piece of the self-energy at the second step, one then obtains Eq.(%
\ref{Tremblay:Self}) above or equivalently Eq.(3) of Ref.\cite{Tremblay:Moukouri:2000,Tremblay:Allen:2003} by considering both longitudinal and transverse
channels and requiring crossing symmetry of the fully-reducible vertex in
the two particle-hole channels as well as consistency with the sum-rule $%
\mathrm{Tr}\left( \Sigma ^{\left( 2\right) }G^{\left( 1\right) }\right) =$ $%
2U\left\langle n_{\uparrow }n_{\downarrow }\right\rangle $ \cite{Tremblay:Vilk:1997}.
The off-diagonal piece of the exact self-energy Eq.(\ref{Tremblay:self-exact}) on the other hand, reads%
\begin{eqnarray}
\Sigma _{12}^{\left( 2\right) } &=&-UG_{11}^{\left( 1\right) }\left( 4,%
\overline{3}\right) \frac{\delta \Sigma _{12}^{\left( 1\right) }\left(
\overline{3},5\right) }{\delta \phi _{\downarrow }\left( 4^{+},4\right) }
\label{Tremblay:Sigma(2)12} \\
&&+UG_{12}^{\left( 1\right) }\left( 4,4^{+}\right) \delta \left( 4-5\right)
-UG_{12}^{\left( 1\right) }\left( 4,\overline{3}\right) \frac{\delta \Sigma
_{22}^{\left( 1\right) }\left( \overline{3},5\right) }{\delta \phi
_{\downarrow }\left( 4^{+},4\right) }.  \notag
\end{eqnarray}%
The pairing susceptibility mediated by spin fluctuations may now be computed
from the derivative with respect to the source field%
\begin{eqnarray}
\left. \frac{\delta G_{12}^{\left( 2\right) }\left( 1,3\right) }{\delta
\theta \left( 2,4\right) }\right\vert _{\theta =0} &=&G_{11}^{\left(
2\right) }\left( 1,2\right) G_{22}^{\left( 2\right) }\left( 4,3\right)  \\
&&+G_{11}^{\left( 2\right) }\left( 1,\overline{4}\right) \left. \frac{\delta
\Sigma _{12}^{\left( 2\right) }\left( \overline{4},\overline{5}\right) }{%
\delta G_{12}^{\left( 1\right) }\left( \overline{6},\overline{7}\right) }%
\right\vert _{\theta =0}\left. \frac{\delta G_{12}^{\left( 1\right) }\left(
\overline{6},\overline{7}\right) }{\delta \theta \left( 2,4\right) }%
\right\vert _{\theta =0}G_{22}^{\left( 2\right) }\left( \overline{5}%
,3\right)   \notag
\end{eqnarray}%
with the irreducible vertex $\delta \Sigma _{12}^{\left( 2\right) }/\delta
G_{12}^{\left( 1\right) }$ obtained by functional differentiation of Eq.(\ref{Tremblay:Sigma(2)12}). Neglecting $\delta /\delta \phi _{\downarrow }\left( \delta
\Sigma _{12}^{\left( 1\right) }/\delta G_{12}^{\left( 1\right) }\right) ,$
which represents the influence of spin fluctuations on the local piece of
the irreducible vertex, and including the transverse component, we find for
the d-wave pair susceptibility, $\chi _{d}$, the expression that appears in
Eq.(1) of Ref. \cite{Tremblay:Kyung:2003}. The TPSC expression for the pair
susceptibility $\chi _{d}$ contains the bubble part and the first term of
what would be an infinite series if $\Sigma _{12}^{\left( 2\right) }$ in the
irreducible vertex could be differentiated with respect to $G_{12}^{\left(
2\right) }$ instead of $G_{12}^{\left( 1\right) }.$
\end{description}

\begin{figure}[tbp]
\centering  \includegraphics*[width=.7\textwidth]{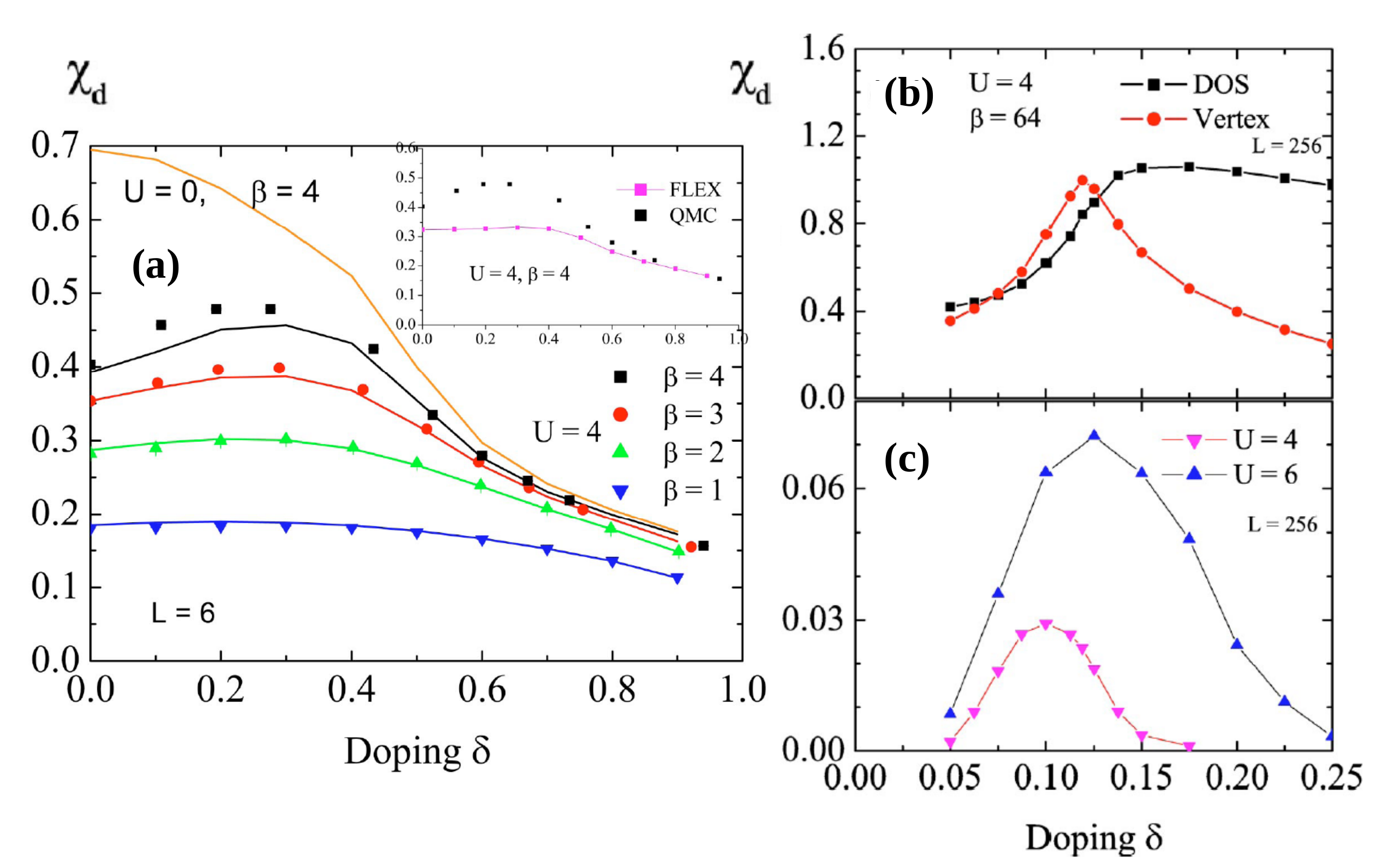}
\caption[Comparisons between QMC and TPSC for the d-wave susceptibility, and estimates of $T_c$]{(Color online) From Ref.\cite{Tremblay:Kyung:2003}. (a) Comparisons between the $d_{x^2-y^2}$ susceptibility
obtained from QMC simulations and from the approach described in the present work. QMC error bars are smaller than the symbols. Analytical results are joined by solid lines. Both calculations are for $U=4$, a $6\times6$ lattice, and four different
temperatures. The case $U=0$, $\beta=4$ is shown for reference.
The size dependence of the results is small at these temperatures. The inset compares QMC and FLEX at $U=4$, $\beta=4$. (b) Contributions from the bubble (DOS) represented by squares and vertex represented by circles. (c) Estimate of $T_c$ using the Thouless criterion for $U=4$ and $U=6$,
$t'=t''=0$.}
\label{Tremblay:f-d-wave}
\end{figure}

Why should we trust the results for the d-wave susceptibility obtained for
TPSC? Let us look again at benchmarks. Fig. \ref{Tremblay:f-d-wave}(a) displays the
d-wave susceptibility obtained from QMC calculations shown as symbols and
from TPSC as lines. Because of the sign problem, it is not practical to do
the QMC calculations at lower temperatures. Nevertheless, the temperatures
are low enough that we see a non-trivial effect, the appearance of a maximum
in susceptibility at finite doping and a substantial increase with
decreasing temperatures. The agreement between QMC and TPSC is to within a
few percent and improves for lower values of $U$. When the interaction
strength reaches the intermediate coupling regime, $U=6$, deviations of the
order of $20\%$ to $30\%$ may occur but the qualitative dependence on
temperature and doping remains accurate. In TPSC the pseudogap is the key
ingredient that leads to a decrease in $\chi _{d}$ in the underdoped regime.
This is easy to understand since the pseudogap leaves fewer states for
pairing at the Fermi level. Another way to say this is that the strong
inelastic scattering that leads to the pseudogap is pair breaking. The inset
shows that previous spin-fluctuation calculations (FLEX) \index{fluctuation exchange approximation} in two dimensions\
\cite{Tremblay:Bickers_dwave:1989,Tremblay:Pao:1995} deviate both qualitatively and
quantitatively from the QMC results. More specifically, in the FLEX approach
$\chi _{d}$ does not show a pronounced maximum at finite doping. This is
because, as we have shown in Fig.\ref{Tremblay:f_QMC_TPSC_FLEX}, in FLEX there is no
pseudogap in the single-particle spectral weight at the Fermi surface\ \cite{Tremblay:Deisz:1996,Tremblay:Moukouri:2000}.

\index{superconductivity!$T_c$ in TPSC}
In two-dimensions, superconductivity is in the
Kosterlitz-Thouless universality class. Vortex physics that is absent in
TPSC is important to understand the precise value of the transition
temperature. Nevertheless, a necessary condition for this transition to
occur is that there is a higher temperature at which pairs form, a sort of
mean-field $T_{c}.$ That $T_{c}$ can be obtained from the so-called Thouless
criterion, i.e. from the temperature at which the d-wave susceptibility
diverges. This divergence occurs because of growing vertex corrections.
Since TPSC contains only the first term in what would be an infinite ladder
sum, we take $T_{c}$ as the temperature at which the bubble (that we call
DOS) and first term of the series (that we call vertex) become equal. This
is illustrated in Fig. \ref{Tremblay:f-d-wave}(b). At $\beta =1/64,$ the doping range
between the intersection of the two curves is below $T_{c}.$ The resulting $%
T_{c}$ versus doping is shown in Fig. \ref{Tremblay:f-d-wave}(c). In that
calculation, $\left\langle n_{\uparrow }n_{\downarrow }\right\rangle $ is
fixed at its value at $T^{\ast }.$

\index{renormalization group approaches!and TPSC}
\index{superconductivity}
There is clearly an additional approximation in finding $T_{c}$ with TPSC
that goes beyond what can be checked with QMC calculations. How can we be
sure that this is correct? First there is consistency with other
weak-coupling approaches. For example, Ref.\cite{Tremblay:Hankevych:2003} has
shown consistency with cutoff renormalization group technique
\cite{Tremblay:Honerkamp:2001} for competing ferromagnetic, antiferromagnetic and d-wave
superconductivity. Second, there is consistency as well with Quantum Cluster
Approaches that are best at strong-coupling. Indeed, an extensive study as a
function of system size by the group of Jarrell \cite{Tremblay:maier_d:2005} has shown
that for $10\%$ doping, $T_{c}$ is in the range $0.02,$ not far from $0.03$
that can be read from Fig. \ref{Tremblay:f-d-wave}(c).

\index{superconductivity!in the Hubbard model}
One of the major theoretical questions in the field of high-temperature
superconductivity has been, ``Is there d-wave superconductivity in the
two-dimensional Hubbard model''? TPSC has contributed to answer this
question. One of the most discouraging early results was that QMC
simulations showed that the pairing susceptibility was smaller at finite $U$
than at $U=0.$ This is clearly seen in Fig.\ \ref{Tremblay:f-d-wave}(a). TPSC allows
us to understand why. At $\beta =4,$ the bubble largely dominates and the
effect can only be pair breaking because of the inelastic scattering. The
vertex, representing exchange of antiferromagnetic fluctuations analogous to
the phonons in ordinary BCS theory, contributes $22\%$ at most at zero doping and much less
at larger doping.
Nevertheless, it clearly increases
the pair susceptibility to bring it in closer agreement with QMC. TPSC allows
us to do the calculation at temperatures much lower than QMC and to verify
that indeed the vertex eventually grows large enough to lead to a
transition. We understand also that in this parameter range the dome shape
comes from the fact that antiferromagnetic fluctuations can both increase
pairing through the vertex and be detrimental through the pseudogap produced
by the large self-energy. Antiferromagnetic fluctuations can both help and
hinder d-wave superconductivity.

Another question is whether the presence of a quantum critical point
below the maximum of the superconducting dome plays a role in
superconductivity. In the case we discussed above, long-range
antiferromagnetic order appears at $T=0$ (not at finite $T$ because of
Mermin-Wagner) up to doping $\delta =0.17$ for $U=4$ \cite{Tremblay:Kyung:2003} and $%
\delta =0.205$ for $U=6$ \cite{Tremblay:Bergeron:2011}. In this case, then, according
to Fig. \ref{Tremblay:f-d-wave}(c) the quantum critical point is far to the right of
the maximum $T_{c}$ but superconductivity can exist to the right of that
point, the more so when $U$ is larger.

How general are the above results? This and many more questions on the
conditions for magnetically mediated superconductivity were studied with
TPSC on the square lattice \textit{at half-filling} for second-neighbor hopping $%
t^{\prime }$ different from zero \cite{Tremblay:Hassan:2008}. At $t^{\prime }=0$ at
half-filling, there is a pseudogap on the whole Fermi surface because of
perfect nesting, so $T_{c}$ vanishes. When $t^{\prime }$ is increased from
zero, the pseudogap is not complete at half-filling and $T_{c}$ is different
from zero. In addition, for $t^{\prime }$ larger than $0.71,$ the Fermi
surface topology changes and the dominant magnetic fluctuations are near $%
\left( 0,\pm \pi \right) ,\left( \pm \pi ,0\right) .$

Additional conclusions of the TPSC study of Ref. \cite{Tremblay:Hassan:2008} as a
function of $t^{\prime }$ and $U$ at half-filling are as follows. First some
qualitative conclusions that could be found from just the BCS gap equation
with an interaction potential given by the static component of the spin
susceptibility \cite{Tremblay:Scalapino:1995}: the symmetry of the d-wave order parameter is
determined by the wave vector of the magnetic fluctuations. Those that are
near $(\pi ,\pi )$ lead to $d_{x^{2}-y^{2}}$-wave ($B_{1g}$)
superconductivity while those that are near $(0,\pi )$ induce $d_{xy}$-wave (%
$B_{2g}$) superconductivity. The dominant wave vector for magnetic
fluctuations is determined by the shape of the Fermi surface so $%
d_{x^{2}-y^{2}}$-wave superconductivity occurs for values of $t^{\prime }$
that are relatively small while $d_{xy}$-wave superconductivity occurs for $%
t^{\prime }>1$. Second, the maximum value that $T_{c}$ can take as a function of $%
t^{\prime }$ increases with interaction strength. TPSC cannot reach the strong-coupling regime where $T_c$ should decrease with $U$.

\index{renormalized classical regime}
One also finds \cite{Tremblay:Hassan:2008} that, contrary to what is expected from BCS, the non-interacting single-particle
density of states does not play a dominant role. At small $t^{\prime }$, $T_{c}$ is reduced by
self-energy effects as discussed above and for intermediate values of $%
t^{\prime }$ the magnetic fluctuations are smaller and incommensurate so no
singlet superconductivity appears. Hence at fixed $U$, \textit{there is an optimal value
of }$t^{\prime }$\textit{\ (frustration)} for superconductivity. For $%
d_{x^{2}-y^{2}}$ superconductivity in under-frustrated systems (small $%
t^{\prime }$) $T_{c}$ occurs below the temperature $T_{X}$ where the
crossover to the renormalized classical regime occurs. In other words, in
under-frustrated systems at $T_{c}$ the antiferromagnetic correlation length
is much larger than the thermal de Broglie wave length and the renormalized
classical spin fluctuations dominate. The opposite relationship between
these lengths occurs for over-frustrated systems ($t^{\prime }$ larger than
optimal) where $T_{c}$ is larger than $T_{X}$ and hence occurs in a regime
where renormalized classical fluctuations do not dominate. The two
temperatures, $T_{c}$ and $T_{X}$, are comparable for optimally frustrated
systems. In all cases, at $T_{c}$ the antiferromagnetic correlation length
is larger than the lattice spacing.

Superconductivity induced by antiferromagnetic fluctuations in weak to
intermediate coupling has also been studied by Moriya and Ueda \cite{Tremblay:Moriya:2003} with the
self-consistent renormalized approach that also satisfies the Mermin-Wagner
theorem. However, in that approach there are adjustable parameters and no
guarantee that the Pauli principle is satisfied so one cannot be certain this
is an accurate solution to the Hubbard model.

That there is d-wave superconductivity in the two-dimensional Hubbard model
has by now been seen by a number of different approaches: variational \footnote{See
contribution of M. Randeria in this volume.} \cite{Tremblay:Giamarchi:1991,Tremblay:Paramekanti:2004}, various Quantum Cluster approaches\footnote{See contribution of D. S\'{e}n\'{e}chal in this volume.} \cite{Tremblay:Senechal:2005,Tremblay:maier_d:2005,Tremblay:Aichhorn:2006,Tremblay:Aichhorn:2007,Tremblay:Haule:2007,Tremblay:kancharla:2008}
functional renormalization group \cite{Tremblay:Honerkamp:2003}, and even at asymptotically
small $U$ by renormalization group \cite{Tremblay:Raghu:2010}. The
retardation that can be observed even tells us that spin fluctuations remain
important for d-wave superconductivity even at strong coupling \cite{Tremblay:Maier:2008,Tremblay:Kyung:2009,Tremblay:Hanke:2010}. The most serious objection to the existence of d-wave
superconductivity comes from a variational and a gaussian Quantum Monte Carlo
approach in Ref.\cite{Tremblay:Aimi:2007}. It could be that d-wave superconductivity in the
two-dimensional Hubbard model is not the absolute minimum but only a local
one. If this were the case, one could conclude that a small interaction term is missing in the Hubbard model to make the d-wave state the ground state. All other studies show that the physical properties of that state are very close the those of actual materials.

\section{More insights on the repulsive model\label{Tremblay:Repulsive}}

The following two sections of this Chapter give a short summary of other results
obtained with TPSC. The purpose is to show what has already been done,
leading to the last section that contains a few open problems that could
possibly be treated with TPSC. These two sections have some of the flavor of
a review article, not that of a pedagogical introduction. In addition, an
important aspect of a real review article is missing: there are very few
references to the rest of the literature on any given topic. For anyone
interested in pursuing some of these problems, the citation index is
highly recommended.

\subsection{Critical behavior and phase transitions\label{Tremblay:critical behavior}}

\index{Pauli principle!and Moriya}
The self-consistent renormalized approach of Moriya-Lonzarich-Taillefer
\cite{Tremblay:Moriya:1985,Tremblay:Lonzarich:1985} was one of the first ones to treat the Hubbard model in two
dimensions in a way that satisfies the Mermin-Wagner theorem. Other
approaches exist for the half-filling case: Schwinger bosons \cite{Tremblay:Arovas:1988} or
constrained spin-waves \cite{Tremblay:Takahashi:1987}. The drawback of Moriya's approach is
that it contains several fitting parameters. Kanamori screening, discussed in Sect.\ref{Tremblay:Mermin
Wagner}, is put by hand, as is the value of the mode coupling constant. In
addition, nothing guarantees the Pauli principle. In other words, Moriya's
approach has much of the same physics as TPSC but it cannot be considered
an accurate solution to the Hubbard model. There is also no prescription to
compute the self-energy in a way that is consistent with double occupancy.

\index{renormalized classical regime}
More generally, the question that arises with TPSC is whether it predicts the
correct universality class. It was shown in Ref. \cite{Tremblay:Dare:1996} that its results are in the
universality class of the spherical model, namely $O\left( N=\infty \right) $
instead of $O\left( N=3\right) $ as it should be for the Hubbard model with
spin-rotation invariance. This result is not surprising since the
self-consistency condition on double occupancy found from the local moment
sum rule Eq.\ (\ref{Tremblay:TPSC_spin}) is very similar to the self-consistency
condition for the spherical model. With the standard convention for critical
exponents, one finds $\gamma /\nu =2,z=2,$ and for dimension $d$ such that
the condition $2<d<4$ is satisfied, we find $\nu =1/\left( d-2\right) .$ This
gives in $d=3,\nu =1,\gamma =2,\beta =1/2,\eta =0$ and $\delta =5.$ This
should be compared with numerical results \cite{Tremblay:Pfeuty:1977} for the 3D
Heisenberg $\left( n=3\right) $ model, $\nu =0.7$ and $\gamma =1.4.$
Clearly, too close to the critical point, or too deep in the renormalized
classical regime in $d=2,$ TPSC looses its accuracy. Results in three
 dimensions can be found in Refs.\cite{Tremblay:Albinet:2000,Tremblay:Arita:2000}.

\subsubsection{Crossover to 3d\label{Tremblay:crossover} \protect\cite{Tremblay:Dare:1996}}

\index{crossover to 3d!antiferromagnetic fluctuations}
The crossover from two- to three-dimensional critical behavior of nearly
antiferromagnetic itinerant electrons was also studied in a regime where the
interplane single-particle motion of electrons is quantum mechanically
incoherent because of thermal fluctuations. The universal renormalized
classical crossover function from $d=2$ to $d=3$ for the susceptibility has
been explicitly computed, as well as a number of other properties such as
the dependence of the N\'{e}el temperature on the ratio between hopping in
the plane $t_{\shortparallel }$ and the hopping perpendicular to it, $%
t_{\perp },$
\begin{equation}
\frac{1}{T_{N}}\sim \frac{T_{N}^{2}}{U_{mf,c}^{2}}\left\vert \ln \left(
\frac{t_{\shortparallel }}{t_{\perp }}\right) \right\vert
\end{equation}%
with $U_{mf,c}\equiv 2/\chi \left( \mathbf{Q}_{d=2},0\right) $ at the $d=2$
pseudogap temperature \cite{Tremblay:Dare:1996}.

\subsubsection{Quantum critical behavior \label{Tremblay:Quantum critical}}

\index{quantum critical behavior}
At $T=0,$ half-filling, the ground state has long-range antiferromagnetic
order. As one dopes, the order becomes incommensurate and eventually
disappears at a critical point that is called \textquotedblleft
quantum critical\textquotedblright\ because it occurs at $T=0.$ Such quantum
critical points are common in heavy-fermion systems for example. One of the
surprising things about this critical point is that it affects the
physics at surprisingly large $T.$

\index{renormalized classical regime}
The quantum critical behavior of TPSC in $d=2$ is in the $z=2$ universality
class, like the self-consistent renormalized theory of Moriya \cite{Tremblay:Moriya:1990}.
Like that theory, it includes some of the logarithmic corrections found in
the renormalization group approach \cite{Tremblay:RoschRMP:2007}. In addition, TPSC can be
quantitative and answer the question, \textquotedblleft How far in $T$ does
the influence of that point extend?" It was found \cite{Tremblay:Roy:2008,Tremblay:RoyPhD:2007} by
explicit numerical calculations away from the renormalized classical regime
of the $d=2$ Hubbard model that logarithmic corrections are not really
apparent in the range $0.01t<T<t$ and that the maximum static spin
susceptibility in the $(T,n)$-plane obeys quantum critical scaling. However,
near the commensurate-incommensurate crossover, one finds obvious
non-universal $T$ and filling $n$ dependence. Everywhere else, the $(T,n)$%
-dependence of the non-universal scale factors is relatively weak. Strong
deviations from scaling occur at $T$ of order $t$, the degeneracy
temperature. That high temperature limit should be contrasted with $J/2$
found in the strong coupling case \cite{Tremblay:Kopp:2005}. In generic cases the
upper limit $T\sim t$ is well-above room temperature. In experiment however,
the non-universality due to the commensurate-incommensurate crossover may
make the identification of quantum critical scaling difficult. In addition,
note that properties other than the maximum spin susceptibility may deviate
from quantum critical scaling at a lower temperature \cite{Tremblay:Bergeron:2011}.

\subsection{Longer range interactions\label{Tremblay:Longer range}}

\index{Hubbard model!near-neighbor interaction}
\index{ETPSC|see Exxtended Two-particle-Self-Consistent}
\index{Extended Two-particle-Self-Consistent Approach}
\index{renormalized classical regime}
Suppose one adds nearest-neighbor repulsion $V$ to the Hubbard model. The
TPSC ansatz Eq. (\ref{Tremblay:Ansatz formal}) can be generalized \cite{Tremblay:Davoudi:2006,Tremblay:Davoudi:2007}. Then one needs to
compute the effect of functional derivatives of the pair-correlation
functions that appear, as in Sect. \ref{Tremblay:TPSC first step}, in the calculation
of spin and charge irreducible vertices. Since the Pauli principle and local
spin and charge sum rules do not suffice, the functional derivatives are
evaluated assuming particle-hole symmetry, which remains approximately true
when the physics is dominated by states close to the Fermi surface. The
resulting theory, called ETPSC, for Extended TPSC, satisfies conservation
laws and the Mermin-Wagner theorem and is in agreement with benchmark
quantum Monte Carlo results. This approach allows to reliably determine the
crossover temperatures toward renormalized-classical regimes, and hence, the
dominant instability as a function of $U$ and $V$. Contrary to RPA, even the
spin fluctuations are modified by the presence of $V.$ Phase diagrams have
been calculated. In the presence of $V$, charge order will generally compete
with spin order \cite{Tremblay:Davoudi:2006,Tremblay:Davoudi:2007}.

\subsection{Frustration \label{Tremblay:Frustration}}

\index{frustration}
\index{Hubbard model!and frustration}
The ETPSC formalism outlined in the previous section is particularly
important to treat interesting problems such as that of the sodium
cobaltates. These compounds are often modeled in an over-simplified way by the two-dimensional Hubbard model on the triangular lattice. To account for charge fluctuations, one must also include nearest-neighbor repulsion $V.$ Even with this
complication this is an oversimplified model.

\index{renormalized classical regime}
The density- and interaction-dependent crossover diagram for spin- and
charge-density wave instabilities of the normal state at arbitrary wave
vector has been computed \cite{Tremblay:Davoudi:2008}. When $U$ dominates over $V$
and electron filling is large, instabilities are mostly in the spin sector
and are controlled by Fermi surface properties. Increasing $V$ eventually
leads to charge instabilities where it is mostly the wave vector dependence
of the vertex that determines the wave vector of the instability rather than
Fermi surface properties. At small filling, nontrivial instabilities appear
only beyond the weak coupling limit. Charge-density wave instabilities are
favored over a wide range of dopings by large $V$ at wave vectors
corresponding to $\sqrt{3}$ $\times \sqrt{3}$ superlattice in real space.
Commensurate fillings do not play a special role for this instability.
Increasing $U$ leads to competition with ferromagnetism. At negative values
of $U$ or $V$, neglecting superconducting fluctuations, one finds that
charge instabilities are favored. In general, the crossover diagram presents
a rich variety of instabilities. Thermal charge-density wave fluctuations in
the renormalized-classical regime can open a pseudogap in the
single-particle spectral weight, just as spin or superconducting
fluctuations \cite{Tremblay:Davoudi:2008}.

\subsection{Thermodynamics, conserving aspects\label{Tremblay:Thermodynamics}}

\index{$\Phi$ derivability}
\index{fluctuation exchange approximation}
\index{skeleton diagram expansion}
\index{thermodynamical consistency}
Conserving approaches are very popular. FLEX \cite{Tremblay:Bickers:1989,Tremblay:Bickers_dwave:1989} is an example.
These approaches are attractive because they guarantee that if one evaluates
the same physical quantity directly from the Green function or from a
derivative of the free energy, the answer will be identical. All that is
needed for a \textquotedblleft conserving\textquotedblright\ approximation
is that the self-energy be generated from a Luttinger Ward functional \cite{Tremblay:LuttingerWard:1960}
that enters the expression for the free energy. In addition, conservation
laws will be satisfied in transport if irreducible vertices are obtained
from functional derivatives of the self-energy. This gives a so-called $\Phi$ derivable theory. Since in perturbation theory
there is an infinite
number of possible such Luttinger-Ward functionals, depending on which closed
two-particle irreducible diagrams constructed from $G$ and the bare interaction one wishes to keep, the constraint of being conserving is
not a very restrictive one. Conserving approximations do not satisfy the
Pauli principle in general and they sometimes give negative values of double occupancy  \cite{Tremblay:Arita:2000}. Various other limitations of conserving
approaches are discussed in Appendix E of Ref. \cite{Tremblay:Vilk:1997}.

TPSC is not obtained from a functional derivative of a Luttinger Ward
functional. Is this a drawback? We have seen that it satisfies conservation
laws for spin and charge at the first step. The question is whether one can
find a unique free energy that is consistent with the one-particle Green function
and collective modes that TPSC focuses on. This question was addressed in
the MSc \cite{Tremblay:RoyMSc:2002} and PhD thesis \cite{Tremblay:RoyPhD:2007} of S\'ebastien Roy. The results are summarized
below. We conclude with an example of thermodynamic calculation in the context of cold atoms.

\subsubsection{Thermodynamic consistency\label{Tremblay:Conserving}}

\index{thermodynamical consistency}
We should really distinguish conservation laws and
thermodynamic consistency. These two notions are sometimes confused, as outlined in the previous paragraph.
We call an approach thermodynamically consistent when all possible
ways of computing the same thermodynamic quantity give the same result.

Obtaining the self-energy from a functional derivative of the Luttinger-Ward
functional leads to thermodynamic consistency. In TPSC there is a change in perspective.
Instead of looking for an approximation for the free energy and then
deducing everything else consistently, we find a single particle
Green's function $G^{\left( 2\right) },$ or equivalently $\Sigma ^{\left(
2\right) },$ as well as double occupancy
\begin{equation}
\frac{1}{2}\mathrm{Tr}\left( \Sigma ^{\left( 2\right) }G^{\left( 2\right)
}\right) =U\left\langle n_{\downarrow }n_{\uparrow }\right\rangle ^{\left(
2\right) }.  \label{Tremblay:TrSigG(2)}
\end{equation}
and deduce everything else: The free energy from integration and
the irreducible vertices for transport quantities from functional
derivatives of $\Sigma ^{\left( 2\right) }$(Sect.\ \ref{Tremblay:Vertex}).

There are three ways to extract the free energy from integration:
(a) coupling constant integration of double occupancy (b) integration of
$\mu(n)$ and (c)
integration of the specific heat calculated from the total energy. With the free
energy, all thermodynamic quantities can be obtained. If we make sure that
the three ways to compute the free energy give the same result, then there
is thermodynamic consistency.

With the above expression for double occupancy, Eq.(\ref{Tremblay:TrSigG(2)}), the
three different ways of obtaining the thermodynamic quantities are all based
on the same object $G^{\left( 2\right) }.$ If $G^{\left( 2\right) }\;$ were the
exact solution, they would have to be consistent$.$ However $G^{\left(
2\right) }$ is approximate. Since $G^{\left( 2\right) }$ satisfies all the
requirements for a physical Green function, it is likely to be the exact
solution of some Hamiltonian $H$ that is close to, but slightly different from, the Hubbard model. For example, $H$ could have longer range interactions. In
deriving the formulas for the free energy, we assume that we are
working with the Hubbard model. Hence, there is no guarantee that all three
methods of obtaining $F$ will give the same result. One can check this numerically in principle.
A simpler test, admittedly less stringent, is to compare $n\left( T,\mu ,U\right) $ obtained
from derivatives of the three different $F.$'s For the nearest-neighbor
hopping model with $\beta =10,$ $U=4$ for example, the results are identical
in the percent range, except deep in the renormalized classical regime close
to half-filling where TPSC anyway fails \cite{Tremblay:RoyMSc:2002,Tremblay:RoyPhD:2007}.
\index{renormalized classical regime}

The specific heat was calculated for the nearest-neighbor
hopping Hubbard model at half-filling as a function of temperature.  TPSC reproduces
the peak observed at small temperature in QMC \cite{Tremblay:Paiva:2001}. It is associated with the entrance in the
renormalized classical regime. Physically, the low temperature peak is a remnant of
the specific heat jump that would occur at finite temperature in mean-field
theory.

\subsubsection{Cold atoms, entropy\label{Tremblay:Entropy}}

\index{cold atoms}
\index{optical lattices!and entropy}
\index{entropy}
In the context of cold atoms on optical lattices, adiabatic cooling can be
used to reach interesting low $T$ regimes such as the pseudogap or ordered
phases by manipulating the scattering length or the strength of the
laser-induced lattice potential. TPSC has been used \cite{Tremblay:Dare:2007}, and
compared with QMC calculations, to provide isentropic curves for the two-
and three-dimensional Hubbard models at half-filling. Since double occupancy$%
\ D$ is extremely accurate in TPSC, the entropy $S$ was computed by
integrating the Maxwell relation%
\begin{equation}
\left( \frac{\partial S}{\partial U}\right) _{T,n}=-\left( \frac{\partial D}{%
\partial T}\right) _{U,n}
\end{equation}%
with $S\left( T,U=0\right) $ the known constant of integration.

The main findings are that adiabatically turning on the interaction in $d=2$
to cool the system is not very effective. In three dimensions, adiabatic
cooling to the antiferromagnetic phase can be achieved in such a manner,
although the cooling efficiency is not as high as initially suggested by
dynamical mean-field theory \cite{Tremblay:WernerHassan:2005}. Adiabatic cooling by turning off the repulsion beginning at strong coupling is possible in certain cases.

\subsection{Vertex corrections and conservation laws\label{Tremblay:Vertex}}

\index{vertex corrections!for conductivity}
Using the functional derivative formalism of Baym and Kadanoff, \cite{Tremblay:BaymKadanoff:1962} it is
possible to find an expression within TPSC for the optical conductivity that
satisfies conservation laws and hence the f-sum rule. In other words, one
can include vertex corrections. Note that the f-sum rule this times involves
the momentum distribution $n_{\mathbf{k}}^{\left( 2\right) }$ obtained from the best self-energy.

The two types of vertex corrections that are found \cite{Tremblay:Bergeron:2011} are the
antiferromagnetic analogs of the Aslamasov-Larkin and Maki-Thompson
contributions of superconducting fluctuations to the conductivity but,
contrary to the latter, they include non-perturbative effects. The
calculations are impossible unless a number of advanced numerical algorithms
are used. Take the case with nearest-neighbor hopping only.\cite{Tremblay:Bergeron:2011} In the pseudogap regime induced by two-dimensional
antiferromagnetic fluctuations, the effect of vertex corrections is
dramatic. Without vertex corrections the resistivity increases as we enter
the pseudogap regime. Instead, vertex corrections lead to a drop in
resistivity, as observed in a number of high temperature superconductors. At
high temperature, the resistivity naturally saturates at the Ioffe-Regel
limit. At the quantum critical point and beyond, the resistivity displays
both linear and quadratic temperature dependence. The disappearance of
superconductivity in the over-doped regime is correlated with the
disappearance of the linear term in the $T$ dependence of the resistivity
\cite{Tremblay:Cooper:2009,Tremblay:Auban:2009}. The relation to the physics of hot spots and
 results for other band structures ($t^\prime\neq 0$) should appear soon.

\section{Attractive Hubbard model \label{Tremblay:Attractive Hubbard}}

\index{Hubbard model!attractive}
Working in Nambu space and following a formal procedure analogous to that
explained in Sect. \ref{Tremblay:Formal}, one can derive TPSC for the attractive
Hubbard model \cite{Tremblay:Allen:2001,Tremblay:Allen:2003,Tremblay:AllenPhD:2000}. The
irreducible vertex $U_{pp}$ in the particle-particle singlet channel is
given by
\begin{equation}
U_{pp}\left\langle n_{\downarrow }\right\rangle \left\langle \left(
1-n_{\uparrow }\right) \right\rangle =U\left\langle n_{\downarrow }\left(
1-n_{\uparrow }\right) \right\rangle
\end{equation}%
and is determined self-consistently at the two-particle level by the local
pair sum-rule%
\begin{equation}
\left\langle n_{\downarrow }n_{\uparrow }\right\rangle =\langle \Delta
^{\dagger }\Delta \rangle =\frac{T}{N}\sum_{q}\chi _{p}^{\left( 1\right)
}(q)\exp (-i\omega _{n}0^{-})
\end{equation}%
with
\begin{equation}
\chi _{p}^{\left( 1\right) }(q)=\frac{\chi _{0}^{\left( 1\right) }(q)}{%
1+U_{pp}\chi _{0}^{\left( 1\right) }(q)}  \label{Tremblay:Chi_pp}
\end{equation}%
and the irreducible particle-particle susceptibility
\begin{equation}
\chi _{0}^{\left( 1\right) }(q)=\frac{T}{N}\sum_{k}G_{\sigma }^{\left(
1\right) }(q-k)G_{-\sigma }^{\left( 1\right) }(k)\;.
\end{equation}%
Again the Pauli principle and a number of crucial sum rules are satisfied.
So is the Mermin-Wagner theorem.

In the second step of the approximation, an improved expression for the
self-energy is obtained by using the results of the first step in an exact
expression for the self-energy, to obtain,
\begin{equation}
\Sigma _{\sigma }^{\left( 2\right) }(k)=Un_{-\sigma }-U\frac{T}{N}%
\sum_{q}U_{pp}\chi _{p}^{\left( 1\right) }(q)G_{-\sigma }^{\left( 1\right)
}(q-k),  \label{Tremblay:Sigma(2)}
\end{equation}%
where $q=\left( i\omega _{n},\mathbf{q}\right) .$ This is a cooperon-like
formula. The required vertex corrections are included as required by the
absence of a Migdal theorem. Comparisons with other approaches can be found in Ref.\cite{Tremblay:Verga:2005}.

\subsubsection{Pseudogap from superconductivity in attractive Hubbard model}

\index{pseudogap!and superconductivity}
\index{superconductivity!pseudogap}
Using the TPSC for the attractive Hubbard model, quantitative agreement with
Monte Carlo calculations is obtained for both single-particle and
two-particle quantities \cite{Tremblay:Kyung:2001}. As discussed for the
repulsive case in Sect.(\ref{Tremblay:Sec Pseudogap}) one obtains a pseudogap in both the
density of states and the single-particle spectral weight \footnote{For a pseudogap in the single-particle spectral weight, it is important not to assume a Migdal theorem, \cite{Tremblay:Fujimoto:2002,Tremblay:Yanase:2004} and include vertex corrections \cite{Tremblay:Vilk:1997}.} below some
characteristic temperature $T^{\ast }$. It was even checked in QMC
calculations that the ratio of the thermal de Broglie wavelength to the
pairing correlation length must be larger than unity to observe the
pseudogap \cite{Tremblay:Allen:1999}. The pseudogap, also found in Ref.\cite{Tremblay:Rohe:2001} for example, reflects precursors of Bogoliubov
quasiparticles that are not local pairs, contrary to what is often discussed
in the context of the crossover from BCS to Bose Einstein condensation \cite{Tremblay:Randeria:1992}.

With increasing temperature the spectral weight fills in the pseudogap
instead of closing it \cite{Tremblay:Kyung:2001}. This type of behavior is obtained in high-temperature
superconductors. The pseudogap appears earlier in the density of states than
in the spectral function. A characteristic behavior observed at strong
coupling appears already in TPSC at weak to intermediate coupling, namely,
small temperature changes around $T^{\ast }$ can modify the spectral weight
over frequency scales much larger than temperature \cite{Tremblay:Kyung:2001}.

Our earlier discussion about Kosterlitz-Thouless physics in Sect.(\ref{Tremblay:d-wave}) is valid in this case as well. In the attractive Hubbard model,
the superconducting transition temperature has a dome shape because at half-filling the symmetry is SO$\left( 3\right) $ so
the $T_{c}$ there vanishes while it is given by the finite
Kosterlitz-Thouless $T_{c}$  \cite{Tremblay:Moreo:1992} elsewhere. The pseudogap temperature on the
other hand decreases monotonically from half-filling where it is largest.
This exemplifies the fact that symmetry and dimension are important to
understand pseudogap physics at weak to intermediate coupling \cite{Tremblay:Allen:1999}.

\section{Open problems\label{Tremblay:Open}}

At weak to intermediate coupling, TPSC gives the best agreement with
benchmark Quantum Monte Carlo methods. Its strength, compared with all other
methods, resides in a non-perturbative treatment of the Hubbard model that
satisfies the Pauli principle and the Mermin-Wagner theorem, in addition to
a number of other exact constraints. Also, one works in the infinite size
limit so the effect of collective fluctuations is not limited to a
small lattice like in dynamical mean-field theory \cite{Tremblay:Georges:1996} and its generalizations \footnote{See contributions of D. Vollhardt, D. S\'en\'echal,
M. Potthoff and M. Jarrell in this volume.} \cite{Tremblay:Maier:2005,Tremblay:KotliarRMP:2006}.

The main weakness of TPSC is the difficulty to extend the method beyond the
one-band Hubbard model. One needs to find enough sum rules to determine the irreducible vertices.  This can be seen as a challenge and an opportunity for creativity.

For example, to include nearest-neighbor Coulomb repulsion, one needs a way
to evaluate functional derivatives of pair correlation functions to obtain
irreducible vertices. It has been possible achieve this \cite{Tremblay:Davoudi:2007}, as discussed in Sect.(\ref{Tremblay:Longer range})%
, but every new problem is different. As another example,
take the case of more than one band. Then the irreducible vertices become a
matrix in band index and one does not have enough obvious sum rules to
evaluate all the matrix elements \cite{Tremblay:Dare:unpublished}.

In the presence of, say, antiferromagnetism the number of irreducible
vertices also multiplies and one faces the same type of challenge. To treat
long-range ordered states with TPSC, it might be easier to start with
simpler broken symmetries such as ferromagnetism or the Pomeranchuk
instability \cite{Tremblay:HalbothPom:2000}. The interest of treating long range order is clear.
For example, the renormalized classical regime of antiferromagnetic
fluctuations is presently inaccessible if $T$ is much smaller than $T^{\ast
}.$ Starting from the ordered state may offer and alternative \cite{Tremblay:borejsza:2004}.

The question of the interplay of disorder and interactions is a difficult
but topical one. Far from the Anderson disorder-induced metal-insulator
transition, the impurity averaging technique \cite{Tremblay:AGD} may prove a useful way
to introduce disorder in TPSC. One may then answer the question of what
happens to the $\xi >\xi _{th}$ criterion for pseudogap when the mean-free
path becomes shorter than the thermal de Broglie wave length.

Climbing the ladder of difficult problems, the case of strong coupling \cite{Tremblay:Georges:1996,Tremblay:Maier:2005,Tremblay:KotliarRMP:2006} is a
real challenge. At strong coupling the self-energy is singular at small
frequencies. In fact it diverges as $1/\omega $ at half-filling. This is
inconsistent with the starting point of TPSC where the self-energy is
constant. Perhaps there is a way to start from the self-energy in the atomic
limit inspired by methods that
allow for multiple poles to zeroth order \cite{Tremblay:Mancini:2004}, or some other way \cite{Tremblay:Saso:2000},
but it is an unsolved problem for now.

Some problems, by contrast, appear straightforward but they can be very
tedious. For example, in the presence of incommensurate magnetic
fluctuations where singlet d-wave pairing does not occur, can triplet
pairing take over? One can proceed along the lines of the derivation for
d-wave superconductivity \cite{Tremblay:Hassan:2008} but with matrix source fields to generate triplet
pairing in the Nambu formalism. The irreducible pairing vertex would again
be obtained from functional derivatives of a matrix $\Sigma ^{\left(
2\right) }.$

Paring in the attractive Hubbard model is much more straightforward, as we
have seen. It appears at the first step, without the need to generate
irreducible vertices from $\Sigma ^{\left( 2\right) }.$ Nevertheless, to
study the triplet channel in the attractive Hubbard model, one needs to
introduce near-neighbor attraction $V$. That leads to the problems mentioned
above in the repulsive case. As a curiosity, one could also investigate
whether functional derivatives of $\Sigma ^{\left( 2\right) }$ in the
attractive Hubbard model can mediate formation of order in some
particle-hole channel. This might be a first step towards developing a
method to take into account the different channels on the same footing in
TPSC, \cite{Tremblay:Fresard:2000} as is done in renormalization group approaches \cite{Tremblay:HonerkampFRG:2001,Tremblay:HonerkampSalmhofer:2001}. It has been found recently within the renormalization group that
in quasi one-dimensional systems there is a strong interference between
antiferromagnetism and unconventional superconductivity \cite{Tremblay:Sedeki:2009}.

Proceeding along the lines of Ref.\ \cite{Tremblay:Bergeron:2011} for the conductivity, it
is also clearly possible to compute other transport quantities, such as the
thermopower, but it is a serious computational challenge.

There are roadblocks, but there are also opportunities for original
solutions and breakthroughs.

\begin{acknowledgement}
I am indebted to Yury Vilk who was at the origin of most of the ideas on
TPSC and to Bumsoo Kyung who has collaborated on numerous TPSC project. In addition, many students, postdocs and colleagues over the years
have worked extremely hard and made use of their creative powers and originality to extend and apply
this approach to a large number of problems. Students and postdocs include
in chronological order Liang Chen, Alain Veilleux, Anne-Marie Dar\'{e},
Steve Allen, Hugo Touchette, Samuel Moukouri, Bumsoo Kyung, Fran\c{c}ois
Lemay, David Poulin, Jean-S\'{e}bastien Landry, Vasyl Hankevych, Bahman
Davoudi, Syed Raghib Hassan, S\'{e}bastien Roy, Charles Brillon and Dominic Bergeron. I am also indebted to
my colleagues David S\'{e}n\'{e}chal, Claude Bourbonnais, and collaborators Gilbert Albinet and Anne-Marie Dar\'{e}. I am grateful to D. S\'{e}n\'{e}chal and D. Bergeron for critical comments on this work. This work was partially
supported by NSERC (Canada) and by the Tier I Canada Research Chair Program
(A.-M.S.T.).
\end{acknowledgement}